\documentclass[longauth]{aa}  

\usepackage{graphicx}
\usepackage{booktabs}
\usepackage{url}
\usepackage{xcolor}

\usepackage[colorlinks = true,
            linkcolor = blue,
            urlcolor  = blue,
            citecolor = blue,
            anchorcolor = blue]{hyperref}
            
\usepackage{scalerel}
\usepackage{tikz}

\newcommand{\emth}[1]{\ensuremath{#1}\xspace}

\newcommand{\tstar}{TOI-519\xspace}
\newcommand{\tplanet}{TOI-519 b\xspace}
\newcommand{\tic}{TIC~218795833\xspace}
\newcommand{\pytransit}{\textsc{PyTransit}\xspace}
\newcommand{\ldtk}{\textsc{LDTk}\xspace}

\newcommand{\gcm}{\emth{\mathrm{g\,cm^{-3}}}}
\newcommand{\smass}{\emth{\mathrm{M_\star}}}
\newcommand{\srad}{\emth{\mathrm{R_\star}}}
\newcommand{\teff}{\ensuremath{T_\mathrm{eff}}\xspace}

\newcommand{\teffc}{\ensuremath{T_\mathrm{eff,C}}\xspace}

\newcommand{\mjup}{\ensuremath{M_\mathrm{Jup}}\xspace}
\newcommand{\rjup}{\ensuremath{R_\mathrm{Jup}}\xspace}
\newcommand{\msun}{\ensuremath{M_\odot}\xspace}
\newcommand{\rsun}{\ensuremath{R_\odot}\xspace}

\newcommand{\ud}{\ensuremath{\mathrm{d}\xspace}}

\newcommand{\tess}{\textit{TESS}\xspace}
\newcommand{\corot}{\textit{CoRoT}\xspace}
\newcommand{\kepler}{\textit{Kepler}\xspace}

\newcommand{\UP}[1]{\ensuremath{\mathcal{U}(#1)}\xspace}
\newcommand{\NP}[1]{\ensuremath{\mathcal{N}(#1)}\xspace}

\newcommand{\ktmedian}{0.298\xspace}
\newcommand{\ktlower}{0.290\xspace}
\newcommand{\ktupper}{0.315\xspace}
\newcommand{\rtmedian}{1.07\xspace}
\newcommand{\rtlower}{0.66\xspace}
\newcommand{\rtupper}{1.20\xspace}

\usepackage{academicons}
\usepackage{xcolor}

\newcommand{\orcidicon}[1]{
  \href{https://orcid.org/#1}{\includegraphics[scale=0.18]{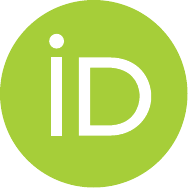}}
}

\usepackage{hyperref}

\usepackage[normalem]{ulem}

\newcommand{\change}[1]{{#1}}

\begin{document} 
	
	\title{\tplanet: a short-period substellar object around an M dwarf validated using multicolour photometry and phase curve analysis}
	\titlerunning{MuSCAT2 validation of \tplanet}

	\author{H. Parviainen\inst{\ref{iiac},\ref{iull}}\orcidicon{0000-0001-5519-1391}
		\and E. Palle\inst{\ref{iiac},\ref{iull}}\orcidicon{0000-0003-0987-1593}
		\and M.R.~Zapatero-Osorio\inst{\ref{csic}}\orcidicon{0000-0001-5664-2852}
		\and G.~Nowak\inst{\ref{iiac},\ref{iull}} 
		\and A.~Fukui\inst{\ref{iuteps}}\orcidicon{0000-0002-4909-5763}
		\and F.~Murgas\inst{\ref{iiac},\ref{iull}}\orcidicon{0000-0001-9087-1245}
		\and N.~Narita\inst{\ref{ikis},\ref{ijsta},\ref{iabc},\ref{inao},\ref{iiac}}\orcidicon{0000-0001-8511-2981}
		\and K.G.~Stassun\inst{\ref{ivandy}}\orcidicon{0000-0002-3481-9052}
		\and J.H.~Livingston\inst{\ref{iutda}}\orcidicon{0000-0002-4881-3620}
		\and K.A.~Collins\inst{\ref{icfa}}\orcidicon{0000-0001-6588-9574}
		\and D.~Hidalgo~Soto\inst{\ref{iiac},\ref{iull}}
		\and V.J.S.~B\'ejar\inst{\ref{iiac},\ref{iull}}
		\and J.~Korth\inst{\ref{ikoln}}\orcidicon{0000-0002-0076-6239}
		\and M.~Monelli\inst{\ref{iiac},\ref{iull}} 
		\and P. Montanes Rodriguez\inst{\ref{iiac},\ref{iull}} 
		%Aphabetical   
		\and N.~Casasayas-Barris\inst{\ref{iiac},\ref{iull}}
		\and G.~Chen\inst{\ref{pmo}}
		\and N.~Crouzet\inst{\ref{iesa}}
		\and J.P.~de Leon\inst{\ref{iutda}}
		\and A.~Hernandez\inst{\ref{iiac},\ref{iull}}
		\and K.~Kawauchi\inst{\ref{iuteps}}
		\and P.~Klagyivik\inst{\ref{iberlin}}
		\and N.~Kusakabe \inst{\ref{iabc},\ref{inao}} 
		\and R.~Luque\inst{\ref{iiac},\ref{iull}}\orcidicon{0000-0002-4671-2957}
		\and M.~Mori\inst{\ref{iutda}}
		\and T.~Nishiumi\inst{\ref{idas},\ref{iabc}}
		\and J.~Prieto-Arranz\inst{\ref{iiac},\ref{iull}}
		\and M.~Tamura\inst{\ref{iutda},\ref{iabc},\ref{inao}}
		\and N.~Watanabe\inst{\ref{inao}}
		% LCO
		\and T.~Gan\inst{\ref{itca}}
		\and K.I.~Collins\inst{\ref{igmu}}
		\and E.L.N.~Jensen\inst{\ref{isc}}
		%\and H.M.~Relles\inst{\ref{ihsca}}
		%- TESS Co-authors
		\and T.~Barclay\inst{\ref{iumary}}\orcidicon{0000-0001-7139-2724}
		%\and D.A.~Caldwell\inst{\ref{iames},\ref{iseti}}\orcidicon{0000-0003-1963-9616}
		\and J.P.~Doty\inst{\ref{inoqsi}}
		\and J.M.~Jenkins\inst{\ref{iames}}\orcidicon{0000-0002-4715-9460}
		\and D.W.~Latham\inst{\ref{icfa}}\orcidicon{0000-0001-9911-7388}
		\and M.~Paegert\inst{\ref{ihsca}}\orcidicon{0000-0001-8120-7457}
		\and G.~Ricker\inst{\ref{imit}}\orcidicon{0000-0003-2058-6662}
		\and D.R.~Rodriguez\inst{\ref{istsci}}\orcidicon{0000-0003-1286-5231}
		\and S.~Seager\inst{\ref{imit},\ref{imitate},\ref{imiteap}}\orcidicon{0000-0002-6892-6948}
		\and A.~Shporer\inst{\ref{imiteap}}\orcidicon{0000-0002-1836-3120}
		\and R.~Vanderspek\inst{\ref{imit}}\orcidicon{0000-0001-6763-6562}
		\and J.~Villase{\~n}or\inst{\ref{imit}}
		\and J.N.~Winn\inst{\ref{iprinceton}}\orcidicon{0000-0002-4265-047X}
		\and B.~Wohler\inst{\ref{iames},\ref{iseti}}\orcidicon{0000-0002-5402-9613}
		\and I.~Wong\inst{\ref{imiteap},\ref{ipegasi}}\orcidicon{0000-0001-9665-8429}
	}

	\institute{
		Instituto de Astrof\'isica de Canarias (IAC), E-38200 La Laguna, Tenerife, Spain\label{iiac}
		\and Dept. Astrof\'isica, Universidad de La Laguna (ULL), E-38206 La Laguna, Tenerife, Spain\label{iull}
		\and Centro de Astrobiologia (CSIC-INTA), Carretera de Ajalvir km 4, 28850 Torrejon de Ardoz, Madrid, Spain\label{csic}
		\and Department of Earth and Planetary Science, The University of Tokyo, Tokyo, Japan \label{iuteps}
		\and Komaba Institute for Science, The University of Tokyo, 3-8-1 Komaba, Meguro, Tokyo 153-8902, Japan \label{ikis}
		\and Japan Science and Technology Agency, PRESTO, 3-8-1 Komaba, Meguro, Tokyo 153-8902, Japan \label{ijsta} 
		\and Astrobiology Center, 2-21-1 Osawa, Mitaka, Tokyo 181-8588, Japan \label{iabc}
		\and National Astronomical Observatory of Japan, 2-21-1 Osawa, Mitaka, Tokyo 181-8588, Japan \label{inao}
        \and Department of Physics and Astronomy, Vanderbilt University, Nashville, TN 37235, USA \label{ivandy}
		\and Department of Astronomy, The University of Tokyo, 7-3-1 Hongo, Bunkyo-ku, Tokyo 113-0033, Japan \label{iutda}
        \and Center for Astrophysics ${\rm \mid}$ Harvard {\rm \&} Smithsonian, 60 Garden Street, Cambridge, MA 02138, USA \label{icfa}
		\and Rheinisches Institut  f\"ur Umweltforschung an der Universit\"at zu K\"oln, Abteilung Planetenforschung, Aachener Str. 209, 50931 K\"oln, Germany \label{ikoln}
		\and  Key Laboratory of Planetary Sciences, Purple Mountain Observatory, Chinese Academy of Sciences, Nanjing 210023, China \label{pmo}
		\and European Space Agency (ESA), European Space Research and Technology Centre (ESTEC), Keplerlaan 1, 2201 AZ Noordwijk, The Netherlands \label{iesa}
		\and Institute of Planetary research, German Aerospace Center, Rutherfordstrasse 2, 12489, Berlin, German \label{iberlin}
        \and Department of Astronomical Science, The Graduated University of Advanced Studies, SOKENDAI, 2-21-1, Osawa, Mitaka, Tokyo, 181-8588, Japan \label{idas}
		\and Department of Astronomy and Tsinghua Centre for Astrophysics, Tsinghua University, Beijing 100084, China \label{itca}
		\and George Mason University, 4400 University Drive, Fairfax, VA, 22030 USA \label{igmu}
		\and Dept.\ of Physics \& Astronomy, Swarthmore College, Swarthmore PA 19081, USA \label{isc}
		\and University of Maryland, Baltimore County, 1000 Hilltop Circle, Baltimore, MD 21250, USA \label{iumary}
		\and NASA Ames Research Center, Moffett Field, CA 94035, USA \label{iames}
		\and SETI Institute, 189 Bernardo Avenue, Suite 100, Mountain View, CA 94043, USA \label{iseti}
		\and Noqsi Aerospace Ltd., 15 Blanchard Avenue, Billerica, MA 01821, USA\label{inoqsi}
		\and Harvard-Smithsonian Center for Astrophysics, 60 Garden St., Cambridge, MA, 02138, USA \label{ihsca}
		\and Department of Physics and Kavli Institute for Astrophysics and Space Research, Massachusetts Institute of Technology, Cambridge, MA 02139, USA \label{imit}
		\and Space Telescope Science Institute, 3700 San Martin Drive, Baltimore, MD 21218, USA \label{istsci}
		\and Department of Aeronautics and Astronautics, Massachusetts Institute of Technology, Cambridge, MA 02139, USA \label{imitate}
		\and Department of Earth, Atmospheric, and Planetary Sciences, Massachusetts Institute of Technology, Cambridge, MA 02139, USA \label{imiteap}
		\and Department of Astrophysical Sciences, Princeton University, Princeton, NJ 08544, USA \label{iprinceton}
		%\and  Department of Physics, Kyoto Sangyo University, Kyoto, Japan \label{ikyo}
		\and 51 Pegasi b Fellow \label{ipegasi}
	}

	\date{Received September 15, 1996; accepted March 16, 1997}
	
	\abstract
	{We report the discovery of \tplanet (\tic), a transiting substellar object ($R = \rtmedian\,\rjup$) orbiting a faint M dwarf
	($V=17.35$) on a 1.26~d orbit. Brown dwarfs and massive planets orbiting M dwarfs on short-period orbits are rare, but more 
	have already been discovered than expected from planet formation models. \tstar is a valuable addition into this group of unlikely
	systems, and adds towards our understanding of the boundaries of planet formation.}
	{We set out to determine the nature of the Transiting Exoplanet Survey Satellite (\tess) object of interest \tplanet.}
	{Our analysis uses a SPOC-pipeline \tess light curve from Sector 7, multicolour transit photometry observed with MuSCAT2 and MuSCAT, 
	and transit photometry observed with the LCOGT telescopes. We estimate the radius of the transiting object using multicolour transit
	modelling, and set upper limits for its mass, effective temperature, and Bond albedo using a phase curve model that includes 
	Doppler boosting, ellipsoidal variations, thermal emission, and reflected light components.}
	{\tplanet is a substellar object with a radius posterior median of \rtmedian~\rjup and 5th and 95th percentiles of \rtlower and 
	\rtupper~\rjup, respectively, where most of the uncertainty comes from the uncertainty in the stellar radius. The phase curve
	analysis sets an upper effective temperature limit of 1800~K, an upper Bond albedo limit of 0.49, and a companion mass upper limit of 
	14~\mjup. The companion radius estimate combined with the \teff and mass limits suggests that the companion is
    more likely a planet than a brown dwarf, but a brown-dwarf scenario is more likely a priori given the lack of known
    massive planets in $\approx 1$~day orbits around M dwarfs with $\teff<3800$~K, and the existence of some (but few) brown dwarfs.}
	{}
	
	\maketitle
	
	\section{Introduction}
	\label{sec:introduction}
	
    Current planet formation models predict a very low probability for a low-mass star to harbour a brown dwarf or a massive planet 
    on a short-period orbit \citep{Mordasini2012}, and M dwarf planet occurrence rate studies based on the \kepler data 
    have corroborated this paucity \citep{Dressing2015}.
    However, against the expectations, a set of such objects have been discovered during the 
    last years. Five brown dwarfs\footnote{TOI~263.01 by \citealt{Parviainen2020}, NGTS-7~A~b by \citealt{Jackman2019}, LP~261-75b by 
    \citealt{Irwin2018}, AD~3116 by \citealt{Gillen2017}, and NLTT~41135~b by \citealt{Irwin2010}.}
    and four gas-giant planets\footnote{Kepler-45b by \citealt{Johnson2012}, HATS-6b by \citealt{Hartman2015}, 
    HATS-71b by \citealt{Bakos2018}, and NGTS-1b by \citealt{Bayliss2018}.} are currently known to orbit M~dwarf hosts cooler 
    than 4000~K with orbital periods smaller than five days. The formation and subsequent evolution of these systems is an open question, 
    as is their actual prevalence. A larger sample is required to find out whether the currently known systems are all rare objects,
    born from random formation accidents, or whether these systems belong to a family with a common formation path.
    
	The \emph{Transiting Exoplanet Survey Satellite} (\tess) \citep{Ricker2014} recently completed the second half of its two-year primary mission, and has discovered to date \change{over} two thousand transiting planet candidates (\emph{TESS objects of interest}, or \emph{TOI}s). However, since various astrophysical phenomena can lead to a photometric 
	signal mimicking an exoplanet transit \citep{Cameron2012}, only a fraction of the candidates are legitimate planets
	\citep{Moutou2009,Almenara2009,Santerne2012,Fressin2013},  and the true nature of each individual candidate needs to be resolved by follow-up
	observations \citep{Cabrera2017a,Mullally2018}. A mass estimate based on radial velocity (RV) measurements is the most reliable 
	way to confirm a planet candidate, but RV observations are practical only for a subset of candidates \citep{Parviainen2019}.
	
	We recently reported the validation of TOI-263.01, a substellar companion orbiting an M~dwarf on an extremely short-period orbit 
	of 0.56 days \citep{Parviainen2020}. The validation was based on ground-based multicolour photometry following a multicolour transit
	modelling approach described in \citet{Parviainen2019}. This approach models transit light curves observed in different passbands 
	(filters) jointly, and yields posterior estimates for the usual quantities of interest (QOIs) in transiting planet light curve analysis, 
	such as orbital period, impact parameter, stellar density, and an estimate for the \emph{true companion radius ratio}. The 
	true radius ratio is a conservative radius ratio estimate\footnote{Here a "conservative radius ratio estimate" means that it should 
	not underestimate the radius ratio, but rather give its reliable upper limit when assuming complete ignorance about the possible
	third light contamination.} that accounts for possible light contamination from unresolved objects 
    inside the photometry aperture. The true radius ratio combined with a stellar radius estimate gives the 
    absolute (conservative) radius of the companion, and if this is securely below the theoretical lower radius limit for a brown dwarf 
    ($\sim 0.8\,\rjup$, \citealt{Burrows2011}), the candidate can be considered a validated planet. If the true radius is $\sim 1\,\rjup$, the nature of
    the companion is ambiguous due to the mass-radius degeneracy for objects with masses in gas giant planet and brown dwarf regime, and for
    radii larger than $1\,\rjup$ the probability that the object is a brown dwarf or a low-mass star increases rapidly.

	We report the discovery of \tplanet, a transiting substellar object ($\rtlower\,\rjup < R < \rtupper\,\rjup$, where the lower and 
	upper limits correspond to 5th and 95th percentiles, respectively) orbiting a faint M dwarf (\tic, see Table~\ref{tbl:star}) on a 
	1.27~d orbit. The object was originally identified in the \tess Sector 7 photometry by the \tess \emph{Science Processing Operations 
	Center} (SPOC) pipeline \citep{Jenkins2016}, and was later followed up from the ground using multicolour transit photometry and
	low-resolution spectroscopy. The planet candidate passes all the SPOC Data Validation tests \citep{Twicken2018}, but the faint host star ($V = 17.35$) makes radial velocity follow up challenging. However,
	the large transit depth makes the system amenable to validation using multicolour transit photometry, although the uncertainties in
	estimating M~dwarf radii complicate the situation by allowing solutions with $R > 1.2\,\rjup$. In this case, a radius estimate is
	not sufficient for validation, and we turn to phase-curve modelling to further constrain the companion's mass and effective temperature.

	\begin{table}[t]    
		\caption{\tstar identifiers, coordinates, properties, and magnitudes. The stellar properties
			are based on a spectrum observed with ALFOSC, \change{and their derivation is described in detail in Sect.~\ref{sec:observations.alfosc_details}}}
		\centering
		\begin{tabular*}{\columnwidth}{@{\extracolsep{\fill}} llrr}
			\toprule\toprule
			\multicolumn{4}{l}{\emph{Main identifiers}}     \\
			\midrule     
			TIC   & & \multicolumn{2}{r}{218795833} \\
			2MASS & & \multicolumn{2}{r}{J08182567-1939465}   \\ 
			WISE  & & \multicolumn{2}{r}{J081825.63-193946.2} \\
			\\
			\multicolumn{4}{l}{\emph{Equatorial coordinates}}     \\
			\midrule            
			RA \,(J2000) &  & \multicolumn{2}{r}{$8^h\,18^m\,25\fs62$}            \\
			Dec (J2000)  &  & \multicolumn{2}{r}{ $-19\degr\,39\arcmin\,46\farcs05$}  \\
			\\     
			\multicolumn{4}{l}{\emph{Stellar parameters }} \\
			\midrule
			Eff. temperature & \teff & [K] & $3350^{+100}_{-200}$\\
			Bolometric flux & $F_\mathrm{bol}$ & [erg\,s$^{-1}$cm$^{-2}$] & $(3.13 \pm 0.11) \times 10^{-11}$ \\
			Mass    & \smass &[\msun]  & $0.369_{-0.097}^{+0.026}$\\
			Radius  & \srad  &[\rsun]  & $ 0.373_{-0.088}^{+0.020} $ \\ 
			Parallax & & [mas] & 8.626\,$\pm$\,0.069\\
			Spectral type & & & M3.5$^{+1.0}_{-0.5}$ \\
			\\
			\multicolumn{4}{l}{\emph{Magnitudes}} \\
			\midrule              
			\centering
			
			Filter & & Magnitude       & Uncertainty  \\
			\midrule     
			TESS & & 14.4347 & 0.0074 \\
			$B$  & & 17.869 & 0.175 \\
			$V$  & & 17.350 & 0.200 \\
			Gaia DR2 & & 15.7067 & 0.0004 \\
			$J$  & & 12.847 & 0.027 \\
			$H$  & & 12.226 & 0.027 \\
			$K$  & & 11.951 & 0.024 \\
			\bottomrule
		\end{tabular*}
		\label{tbl:star}  
	\end{table}

	\section{Observations}
	
	\subsection{\tess photometry}
	\label{sec:observations.tess}
	
	\begin{figure}
		\centering
		\includegraphics[width=\columnwidth]{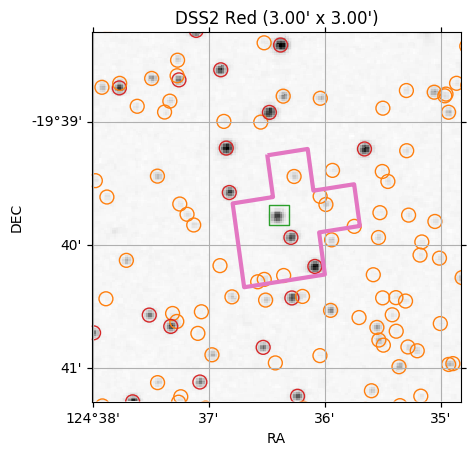}
		\caption{\tstar and its surroundings observed by DSS with the \tess aperture used by the SPOC 
		pipeline shown in pink and \tstar marked with a square. The nearest star lower-right from \tstar 
		introduces a significant amount of flux contamination in the \tstar light curve.}
		\label{fig:field_tess}
	\end{figure}
	
	\tess observed \tplanet during Sectors 7 and 8. Sector 7 was observed for 24.4~days covering 18 transits 
	with a two-minute cadence. Sector 8 was observed for for 24.6 days covering 13 transits (some of the 
	transits occur during useless sections of the light curve), but the two-minute time cadence is not available,
	and the light curve must be created from the Full Frame Image (FFI) data.
	
	We chose to use the Sector 7 Presearch Data Conditioning (PDC) light curve \citep{Stumpe2014,Stumpe2012,Smith2012a}, produced by the SPOC pipeline, 
	for the analysis. We add back the crowding correction ("CROWDSAP")\footnote{The PDC crowding 
	metric, $C$, corresponds to the ratio of the target flux to the total aperture flux, and the contamination 
	defined in this paper, $c$, to the ratio of the
	contaminating flux to the total flux, and the two are related as $c = 1-C$.} removed by the pipeline, 
	since the crowding correction can introduce a bias into our parameter estimation if the crowding is 
	overestimated by the SPOC pipeline.\!\footnote{As is the case here, see Discussion for details.} The \tstar light 
	curve can be expected to contain a significant amount of flux contamination from a nearby star with a similar 
	brightness (see Fig.~\ref{fig:field_tess}), and our parameter estimation approach leads to an independent \tess 
	contamination estimate based on the differences in transit depths between the \tess and ground-based transit
	observations.
	
	The \tess photometry used in the transit analysis consists of 18 3.6~hour-long windows of SPOC data from Sector 7 
	centred around each transit based on the linear ephemeris, and each subset was normalised to its median out-of-transit
	(OOT) level assuming a transit duration of 2.4~h. The photometry has an average point-to-point (ptp) scatter 
	of 18.7~parts per thousand (ppt). We do not detrend the photometry, but model the baseline in the transit analysis. 
	
	The phase curve analysis uses all the SPOC Sector 7 data except the transits. We also created long-cadence light 
	curves for Sectors 7 and 8 using the \textsc{Eleanor} package \citep{Feinstein2019}, since while the short transit 
	duration makes long-cadence less-than-optimal for transit modelling, having two sectors
	of data instead of one could still be beneficial for the phase curve modelling. However, we were not able to
	produce light curves with \textsc{Eleanor} that would have matched the SPOC-produced light curve in quality. 
	The long-cadence light curves show significantly higher systematics that led them to be useless in the phase
	curve analysis.
	
	\subsection{MuSCAT2 photometry}
	\label{sec:observations.muscat2}
\begin{figure}
		\centering
		\includegraphics[width=\columnwidth]{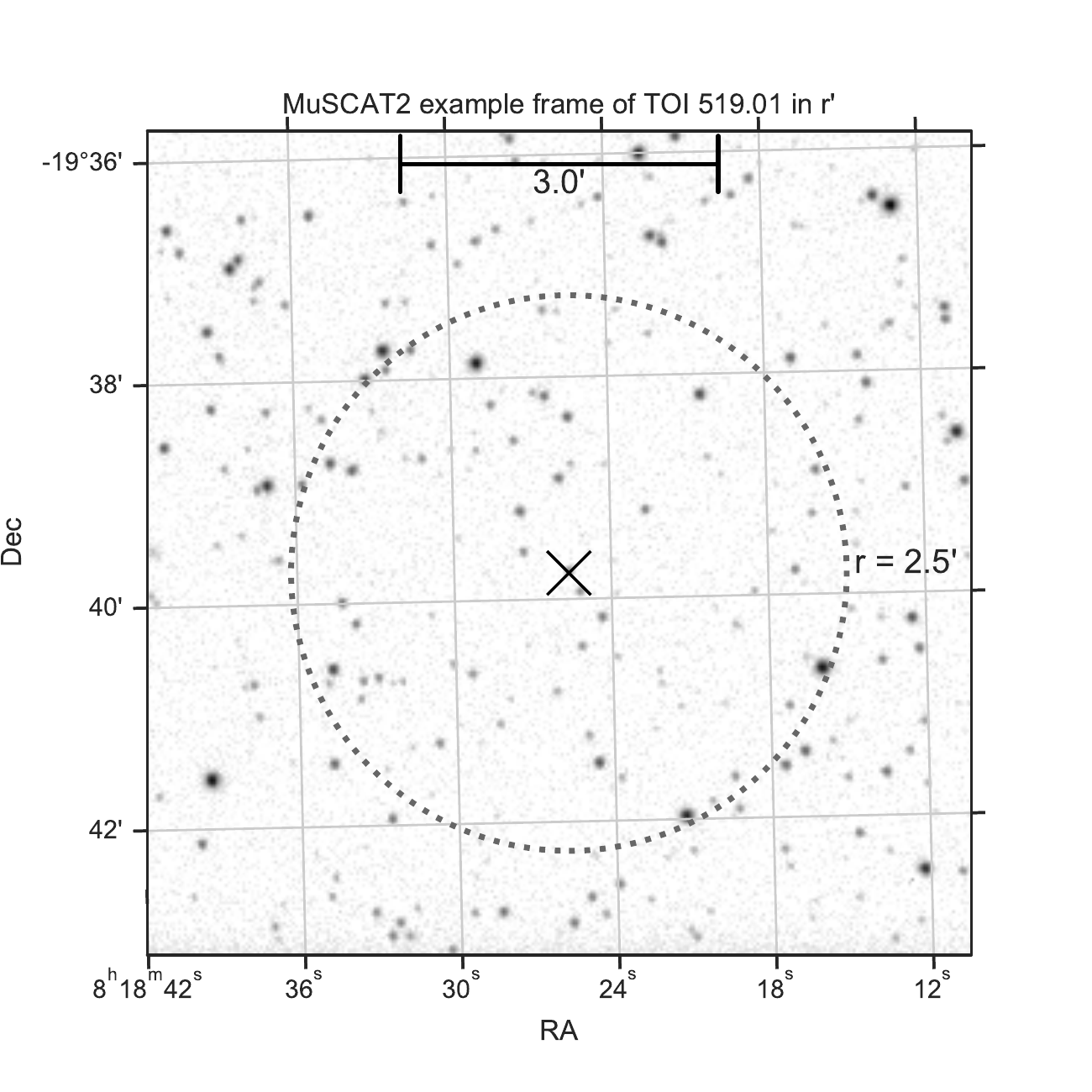}
		\caption{MuSCAT2 field observed in $r$ band. \tstar is marked with a cross, and the dotted circle marks the $2.5\arcmin$-radius region centred around \tstar.}
		\label{fig:m2_field}
	\end{figure}

	We observed four full transits of \tplanet simultaneously in $g$, $r$, $i$, and $z_\mathrm{s}$ bands with the MuSCAT2 
	multicolour imager \citep{Narita2018} installed at the 1.52~m Telescopio Carlos Sanchez (TCS) in the Teide Observatory, Spain,
	on the nights of 22.11.2019, 8.1.2020, 13.1.2020, and 29.2.2020. The exposure times were optimised on per-night
	and per-CCD basis, but were generally between 60 and 90~s. The observing conditions were excellent through
	all the nights (see Fig.~\ref{fig:m2_field} for an example frame). 
	
	The photometry was carried out using standard aperture photometry calibration and reduction steps with a 
	dedicated MuSCAT2 photometry pipeline, as described in \citet{Parviainen2020}. The pipeline calculates
	aperture photometry for the target and a set of comparison stars and aperture sizes, and 
	creates the final relative light curves via global optimisation of a model that aims to find the optimal
	comparison stars and aperture size, while simultaneously modelling the transit and baseline variations
	modelled as a linear combination of a set of covariates.

    \subsection{MuSCAT photometry}
    \label{sec:observations.muscat}
    We also observed one full transit of \tplanet simultaneously in $g$, $r$, and $z_\mathrm{s}$ bands with the multicolour 
    imager MuSCAT \citep{Narita2015} mounted on the 1.88~m telescope at Okayama Astro-Complex on Mt. Chikurinji, Japan, on 30.11.2019. 
    The observation was conducted for 3.4 hours covering the transit, during which the sky condition was excellent. The telescope 
    focus was slightly defocused so that the full width at half maximum (FWHM) of stellar point-spread function (PSF) was 
    around 3\arcsec. The exposure times were set at 60, 40, and 60~s in $g$, $r$, and $z_\mathrm{s}$ bands, respectively.
    
    Image calibration (dark correction and flat fielding) and standard aperture photometry were performed using a custom 
    pipeline \citep{Fukui2011}, with which the aperture size and comparison stars were optimised so that the point-to-point 
    dispersion of the final light curve was minimised. The adopted aperture radius was 10 pixels (3.6\arcsec) for all bands.

	\subsection{LCOGT photometry}
	\label{sec:observations.lco}
	Three full transits of \tplanet were observed using the Las Cumbres Observatory Global Telescope (LCOGT) 1~m network 
	\citep{Brown:2013} in $g$, $i$, and $z_\mathrm{s}$ bands on the nights of 29.03.2019, 01.04.2019, and 16.04.2019, 
	respectively. We used the {\tt TESS Transit Finder}, which is a customised version of the {\tt Tapir} software package 
	\citep{Jensen:2013}, to schedule our transit observations. The $g$ and $z_\mathrm{s}$ transits were observed from the 
	LCOGT node at Cerro Tololo Inter-American Observatory, Chile, and used 60~s and 150~s exposures, respectively. The $i$ 
	transit was observed from the LCOGT node at South Africa Astronomical Observatory, South Africa, and used 150~s exposures. 
	The 1~m telescopes are equipped with $4096\times4096$~pixel LCO SINISTRO cameras having an image scale of 0$\farcs$389 
	pixel$^{-1}$ resulting in a $26\arcmin\times26\arcmin$ field of view. 
	
	The images were calibrated by the standard LCOGT BANZAI pipeline \citep{McCully:2018} and the photometric data were extracted 
	using the {\tt AstroImageJ} ({\tt AIJ}) software package \citep{Collins:2017}. The $g$ and $z_\mathrm{s}$ images have PSFs
	with FWHM $\sim 1\farcs8$, and the $i$ images were defocused resulting in FWHMs $\sim 3\farcs2$. Circular apertures with 
	radius 11, 15, and 10 pixels were used to extract differential photometry in the $g$, $i$, and $z_\mathrm{s}$ bands, 
	respectively. 

	\subsection{Spectroscopy}
	\label{sec:observations.alfosc}
    On 16.3.2020, we obtained the optical low-resolution spectrum
    of \tstar{} with the Alhambra Faint Object Spectrograph and
    Camera (ALFOSC) mounted at the 2.56~m Nordic Optical
    Telescope (NOT) on the Roque de los Muchachos
    Observatory. ALFOSC is equipped with a 2048$\times$2064 CCD
    detector with a pixel scale of 0.2138 \arcsec
    pixel$^{-1}$. We used grism number 5 and an horizontal long
    slit with a width of 1.0\arcsec\!, which yield a nominal
    spectral dispersion of 3.53 \AA~pixel$^{-1}$ and a usable
    wavelength space coverage between 5000 and 9400~\AA. Two
    spectra of 1800 s each were acquired at parallactic angle and
    airmass of 1.51. ALFOSC observations of the white dwarf
    G191--B2B were acquired with the same instrumental setup as
    \tstar{}, with an exposure time of 120 s, and at an airmass
    of 1.65. Raw images were reduced following standard
    procedures at optical wavelengths: bias subtraction,
    flat-fielding using dome flats, and optimal extraction using
    appropriate packages within the IRAF\footnote{Image Reduction
    and Analysis Facility (IRAF) is distributed by the National
    Optical Astronomy Observatories, which are operated by the
    Association of Universities for Research in Astronomy, Inc.,
    under contract with the National Science Foundation.}
    environment. Wavelength calibration was performed with a
    precision of 0.65 \AA~using He\,{\sc i} and Ne\,{\sc i} arc
    lines observed on the same night. The instrumental response
    was corrected using observations of the spectrophotometric
    standard star G191--B2B. Because the primary target and the
    standard star were observed close in time and at a similar
    airmass, we corrected for telluric lines absorption by
    dividing the target data by the spectrum of the standard
    normalised to the continuum. The two individual spectra of
    \tstar{} were combined and the final spectrum, which has a
    spectral resolution of 16 \AA{} ($R \approx 450$ at
    7100~\AA{}), is depicted in Fig.~\ref{fig:toi_519_spectrum}.
    
	\section{Stellar characterisation}
	\label{sec:observations.alfosc_details}
	We used the ALFOSC telluric-free spectrum to determine the spectral type of \tstar{} by measuring various spectroscopic indices, or colour ratios, suitable for M dwarfs. Some of these indices are nearly insensitive to the instrumental correction errors and their sensitivity to low and moderate extinction is also reduced, which makes them reliable indicators of spectral type. Other indices are useful as luminosity and metallicity discriminants. We obtained the flux ratios A, B/A, D/A, and TiO5 defined by \citet{kirk91} and \citet{gizis97}, all of which explore strong atomic lines and molecular bands present in M-type stars. Derived values and a short description of the features covered by the flux ratios are given in Table~\ref{tbl:specind}. All of these indices show very little dispersion in terms of spectral type and indicate that \tstar{} is an M3.0--M3.5 dwarf. This spectral type is fully consistent with the absolute magnitudes of \tstar{} in the optical through mid-infrared wavelengths (see next subsection). 
	 
\begin{figure}
		\centering
		\includegraphics[width=\columnwidth]{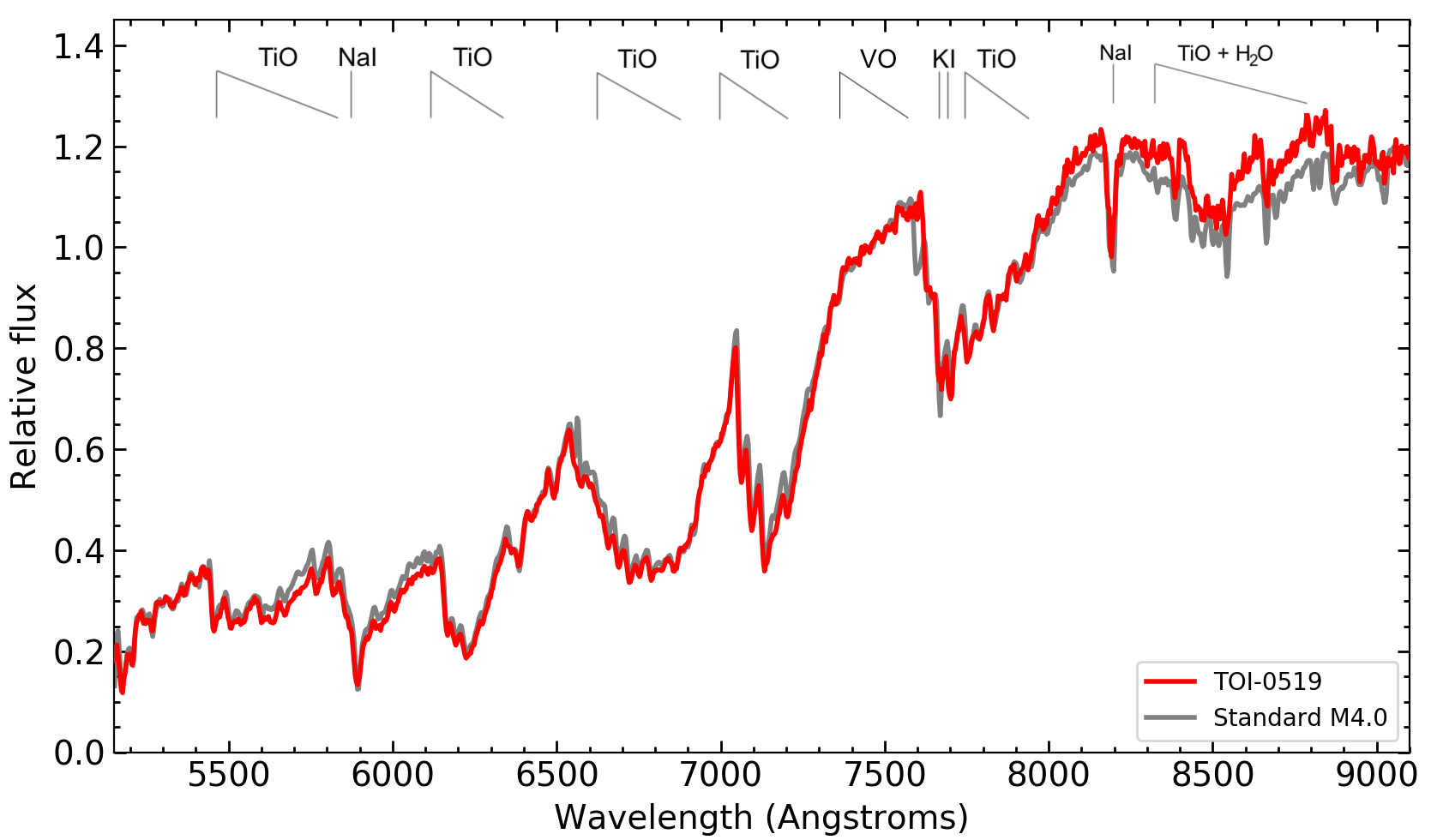}
		\caption{The ALFOSC, telluric-free optical spectrum of \tstar{} is shown in red (spectral resolution of 16 \AA).
		For comparison purposes, the solar-metallicity M4.0 spectral standard template from \citet{Kesseli2017} is also plotted as the grey line (this spectrum has been degraded to the resolution of our target and it is also corrected for the telluric lines absorption). The most significant spectral features are labelled. The spectra are normalised to unity at around 7500 \AA.}
		\label{fig:toi_519_spectrum}
	\end{figure}

	However, spectral indices covering widely separated pseudo-continuum and feature regions yield later spectral types. The PC3 index defined by \citet{martin96}, which measures the spectroscopic slope between two pseudo-continuum points of the optical data, delivers M5 spectral type (Table~\ref{tbl:specind}). The best match to the ALFOSC spectrum among the data set of spectroscopic templates of \citet{Kesseli2017} is provided by M4.0 spectral type as illustrated in Fig.~\ref{fig:toi_519_spectrum}. This discrepancy of about one spectral type may be explained by the presence of a moderate extinction or a higher metallicity. The former scenario, although feasible given the low Galactic latitude of our target ($b \approx +9$ deg), is less likely because of the close distance to \tstar{}. Also, $GJHK$ and {\sl WISE} colours are compatible with one single spectral type, which is a signpost of no or very little extinction towards \tstar{}.  Nevertheless, to explore the high metallicity scenario in detail, higher resolution spectra would be needed. To be conservative, we will adopt a spectral type M3.0--M4.5 for \tstar{}.
	
	From the ALFOSC spectrum, H$\alpha$ is not seen in emission and we can impose a lower limit of 0.5 \AA~to the pseudo-equivalent width of any emission feature around 6563 \AA. Potasium and sodium atomic lines, which are features rather sensitive to temperature and surface gravity, are seen in absorption with strengths similar to those of the M3.0--M4.5 standard stars. This suggests that \tstar{} has a high surface gravity, thus discarding the idea that our target is a very young or a giant star (and the atomic and molecular indices of Table~\ref{tbl:specind} also reject our target being a giant or a subdwarf star). 
	
	From the spectral type---$T_{\rm eff}$ relationship given in \citet{houdebine19}, we derived $T_{\rm eff} = 3350^{+100}_{-200}$ K for \tstar{}, where the quoted
	errors account for spectral types in the interval M3--M4.5. \citet{houdebine19} claimed that their temperature calibration is valid for solar and near-solar
	metallicities (only strongly metal-depleted M dwarfs deviate from this calibration). \tstar{} does not show obvious absorption features due to hydrides (
	e.g., CaH) in the optical spectrum, which indicates that it is not a subdwarf \citep{Kirkpatrick2014}. Using various $T_{\rm eff}$---mass---stellar radii
	relationships available in the literature (e.g., \citealt{Schweitzer2019,houdebine19}; Cifuentes et al$.$, submitted), we obtained that \tstar{} has a radius 
	of $R_* = 0.373^{+0.020}_{-0.088}$ R$_\odot$ and a mass of $M_* = 0.369^{+0.026}_{-0.097}$ M$_\odot$. This mass determination is only slightly larger than that 
	derived from \citet{Mann2019} \change{relations, $M_\star=0.36\pm 0.03\;\msun$}, and both values are consistent at the 1 $\sigma$ level.
	
		\begin{table}[t]    
		\caption{Spectroscopic indices and color ratios.}
		\centering
		\begin{tabular*}{\columnwidth}{@{\extracolsep{\fill}} cccc}
			\toprule\toprule
			\multicolumn{1}{c}{Index} &
			\multicolumn{1}{c}{Feature} &
			\multicolumn{1}{c}{Value} &
			\multicolumn{1}{c}{SpT} \\
			\midrule     
			A    & CaH $\lambda$6975 \AA                      & 1.24 & M3.0 \\
			B/A  & Ti\,{\sc i} $\lambda$7358 \AA              & 0.85 & M3.5 \\
			D/A  & Ca\,{\sc ii} $\lambda$8542 \AA             & 0.87 & M3.5 \\
			TiO5 & TiO $\lambda\lambda$7042--7135 \AA         & 0.48 & M3.0 \\
			PC3  & pseudo-continuum $\lambda\lambda$7569, 8250 \AA & 1.21 & M5.0 \\ 
			\bottomrule
		\end{tabular*}
		\tablefoot{The uncertainty of the indices is 5\%~or less. All spectral types have been rounded to the nearest half subtype.}
		\label{tbl:specind}  
	\end{table}
	
    \subsection{Spectral Energy Distribution}
\begin{figure}[!ht]
	    \centering
	    \includegraphics[width=\columnwidth]{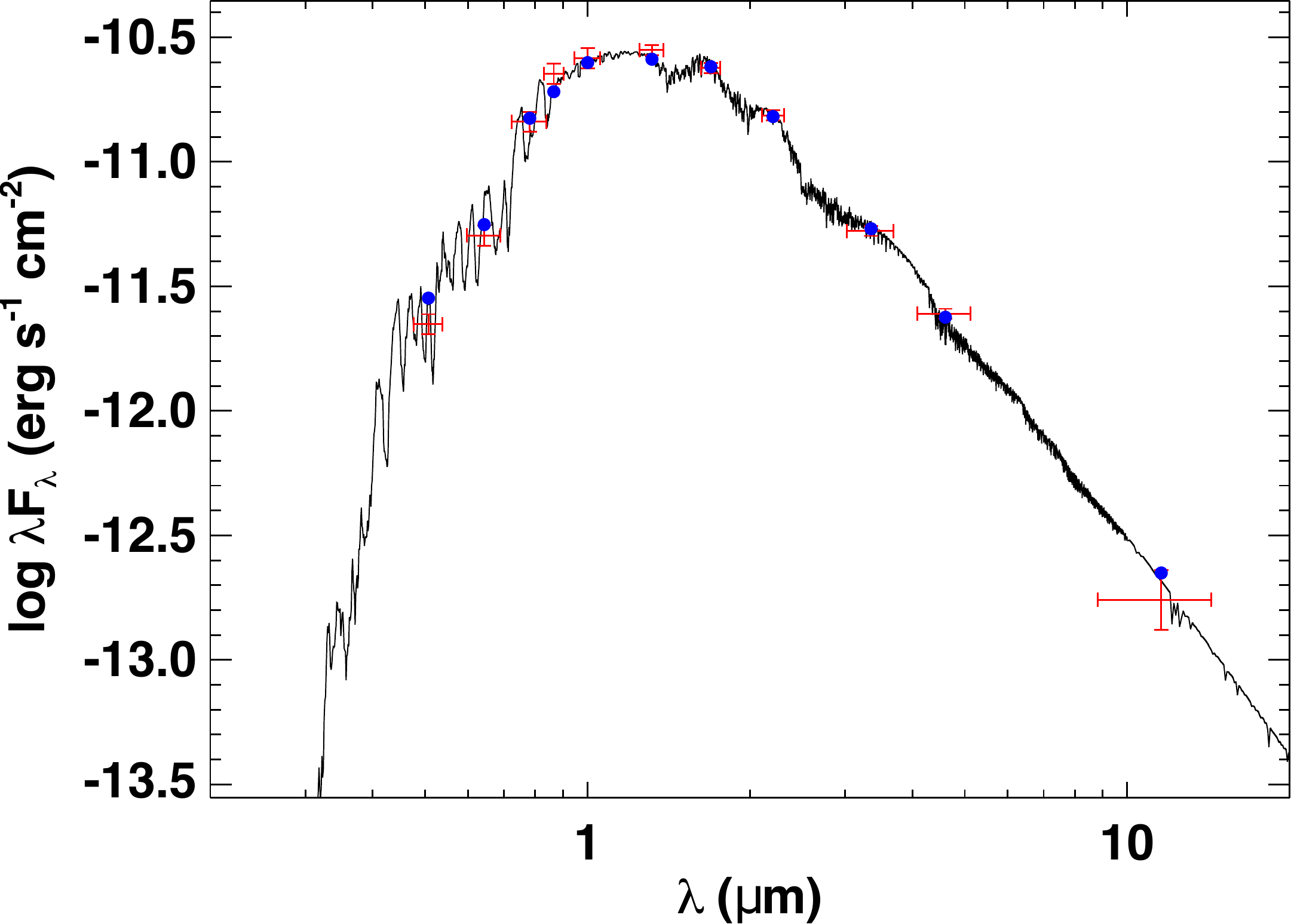}
	\caption{Spectral energy distribution of TOI~519. Red symbols represent the observed photometric measurements, where the horizontal bars represent the effective width of the passband. Blue symbols are the model fluxes from the best-fit NextGen atmosphere model (black).  \label{fig:sed}}
	\end{figure}

    As an independent determination of the stellar parameters, in particular the stellar radius, we performed an analysis of the broadband 
    spectral energy distribution (SED) together with the {\it Gaia\/} DR2 parallax following the procedures described in 
    \citet{Stassun2016} and \citet{Stassun2017a,Stassun2017}. We pulled the $grizy$ magnitudes from the Pan-Starrs database, the $JHK_S$ magnitudes 
    from {\it 2MASS}, and the $W1$--$W3$ magnitudes from {\it WISE}. Together, the available photometry spans the full stellar SED over the 
    wavelength range 0.4--10~$\mu$m (see Fig.~\ref{fig:sed}). 

    We performed a fit using NextGen stellar atmosphere models, adopting the effective temperature from the spectroscopically 
    determined value. The extinction ($A_V$) was set to zero due to the star being very near. The metallicity was left as a free 
    parameter. The resulting fit, shown in Fig.~\ref{fig:sed}, has a reduced $\chi^2$ of 2.5 and a best-fit metallicity of $0.0\pm0.5$. Integrating 
    the model SED gives the bolometric flux at Earth of $F_{\rm bol} =$  $(3.20\pm0.11) \times 10^{-11}$ erg\,s$^{-1}$\,cm$^{-2}$. Taking the 
    $F_{\rm bol}$ and $T_{\rm eff}$ together with the {\it Gaia\/} DR2 parallax, adjusted by $+0.08$~mas to account for the systematic offset 
    reported by \citet{Stassun2018}, gives the stellar radius as $0.342\pm0.031~\rsun$. Finally, we estimate a mass of $0.36\pm0.03$~\msun\ via 
    \citet{Mann2019} empirical M-dwarf relations based on the absolute K-band magnitude. These values are in agreement with the values estimated 
    from the low-resolution spectrum.

	\section{Light curve analysis}
	\label{sec:analysis}
	
	\subsection{Overview}
	\label{sec:analysis.overview}
	
	We modelled the \tess light curves simultaneously with the MuSCAT2, MuSCAT, and LCOGT light curves following
	the approach described in \citet{Parviainen2020} and \citet{Parviainen2019} to characterise the system and obtain
	a robust "true radius ratio" estimate for the companion. Next, we carried out a phase curve analysis using 
	the \tess light curve to constrain the companion's effective temperature and mass. As a double-check, we also carried out 
	separate analyses using only the \tess or the ground-based data, but we do not detail those here. We have assumed 
	zero eccentricity in all the analyses given the short circularisation time scales for a one-day period
	\citep{Dawson2018}.

	The analyses were carried out with a custom Python code based on \pytransit~v2\footnote{\url{https://github.com/hpparvi/pytransit}}
	\citep{Parviainen2015,Parviainen2019}, which includes a physics-based contamination model based 
	on the \textsc{PHOENIX}-calculated stellar spectrum library by \citet{Husser2013}. The limb darkening computations were carried 
	out with \ldtk$\!$\footnote{\url{https://github.com/hpparvi/ldtk}} \citep{Parviainen2015b}, and 
	Markov Chain Monte Carlo (MCMC) sampling was carried out with \textsc{emcee} \citep{Foreman-Mackey2012,Goodman2010}. The code relies on the existing \textsc{Python} packages for scientific computing and astrophysics:
	\textsc{SciPy}, \textsc{NumPy} \citep{VanderWalt2011}, \textsc{AstroPy} \citep{TheAstropyCollaboration2013,Price-Whelan2018}, 
	\textsc{photutils} \citep{Bradley2019}, \textsc{astrometry.net} \citep{Lang2010}, \textsc{IPython} \citep{Perez2007},
	\textsc{Pandas} \citep{Mckinney2010}, \textsc{xarray} \citep{Hoyer2017}, \textsc{matplotlib} \citep{Hunter2007}, and
	\textsc{seaborn}. The analyses are publicly available from GitHub\footnote{\url{https://github.com/hpparvi/parviainen_2020_toi_519}} as Jupyter notebooks.

    \subsection{Multicolour transit analysis}
	The final multicolour photometry dataset consists of the 18 transits in the \tess data from Sector 7, four 
	transits observed simultaneously in four passbands ($g$, $r$, $i$, and $z_\mathrm{s}$) with MuSCAT2, one transit
	observed in three passbands ($g$, $r$, and $z_\mathrm{s}$) with MuSCAT, and two 
	transits observed in two passbands ($i$ and $z$) with the LCOGT telescopes. This sums up to five passbands
	(we consider $z$ and $z_\mathrm{s}$ the same), 25 transits, and 39 light curves.
	
	The analysis follows standard steps for Bayesian parameter estimation \citep{Parviainen2018}. First, 
	we construct a flux model that aims to reproduce the transits and the light curve systematics. 
	Next, we define a noise model to explain the stochastic variability in the observations not explained 
	by the deterministic flux model. Combining the flux model, the noise model, and the observations gives 
	us the likelihood.  Finally, we define the priors on the model parameters, after which we estimate the 
	joint parameter posterior distribution using Markov Chain Monte Carlo (MCMC) sampling.
	
	The posterior estimation begins with a global optimisation run using Differential Evolution
	\citep{Storn1997,Price2005} that results with a 
	population of parameter vectors clumped close to the global posterior mode. This parameter vector 
	population is then used as a starting population for the MCMC sampling with \textsc{emcee}, and the
	sampling is carried out until a suitable posterior sample has been obtained \citep{Parviainen2018}. The model parametrisation, priors, and the construction of the posterior function follow directly \citet{Parviainen2020}.
	
	\subsection{Phase curve analysis}
	\label{sec:eclipse_analysis}
	While multicolour transit analysis allows us to securely estimate the companion's radius ratio, modelling 
	variations in the \tess light curve over its orbital phase gives us a tool to estimate the companion's effective 
	temperature, Bond albedo, and mass \citep{Loeb2003,Mislis2012a,Shporer2017,Shporer2019}. Phase curve modelling is a well-established method for 
	companion mass estimation, and has been widely used to study planets and brown dwarfs found by the \corot and 
	\kepler missions (i.e, CoRoT-3b \citealt{Mazeh2010}; TrES-2b \citealt{Barclay2012}; Kepler-13b \citealt{Shporer2011a} and \citealt{Mislis2012};
	Kepler-91b \citealt{Lillo-Box2014} and \citealt{Barclay2014}; and Kepler-41b \citealt{Quintana2013}, to 
	name a few, and homogeneous phase-curve studies have been also reported by \citealt{Esteves2013,Angerhausen2014}; and \citealt{Esteves2015}).
	
\begin{figure}
		\centering
    	\includegraphics[width=\columnwidth]{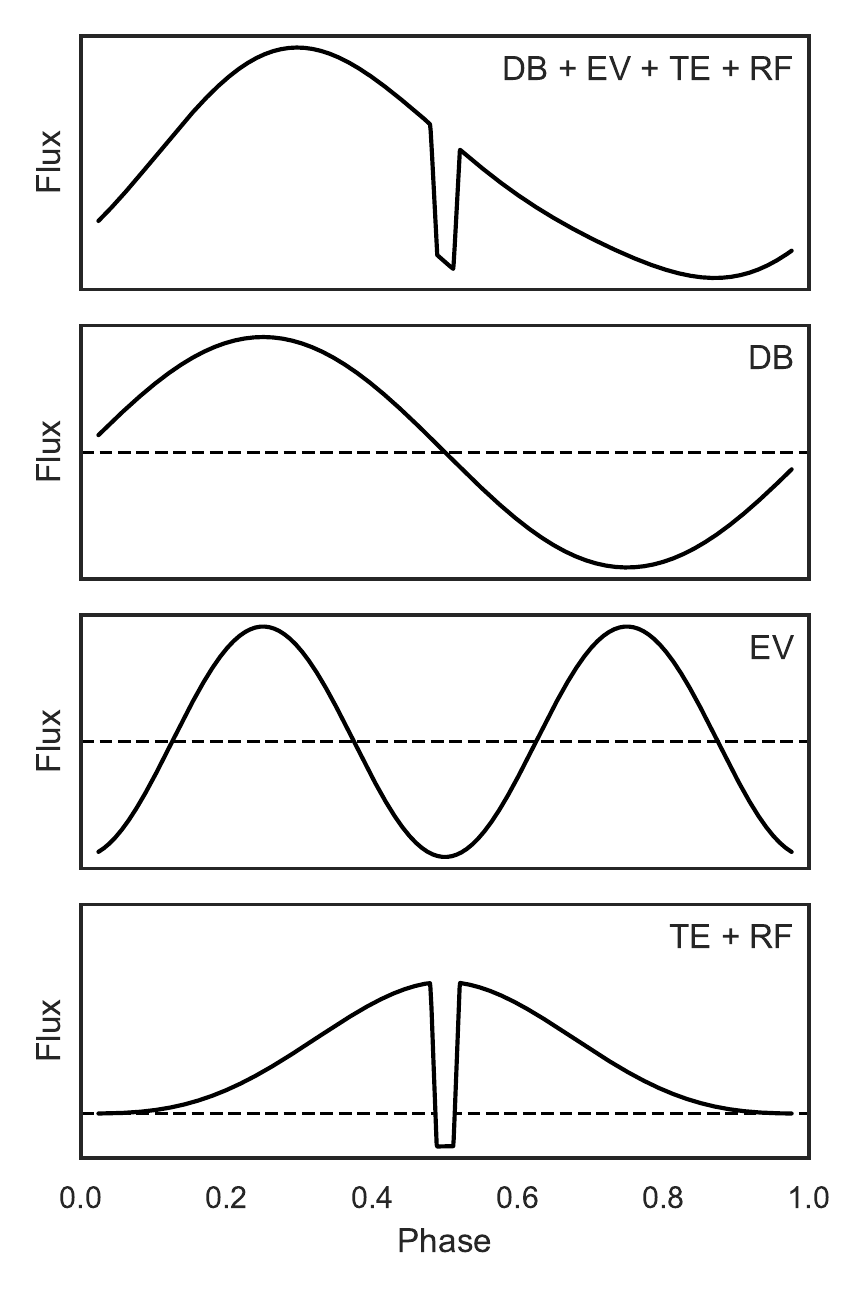}
		\caption{Schematic showing Doppler boosting (DB), ellipsoidal variations
		(EV), thermal emission (TE), and reflection (RF) and all the phase curve components combined
		as a function of orbital phase.}
		\label{fig:phase_curve_schematic}
	\end{figure}

\begin{figure*}
		\centering
    	\includegraphics[width=\textwidth]{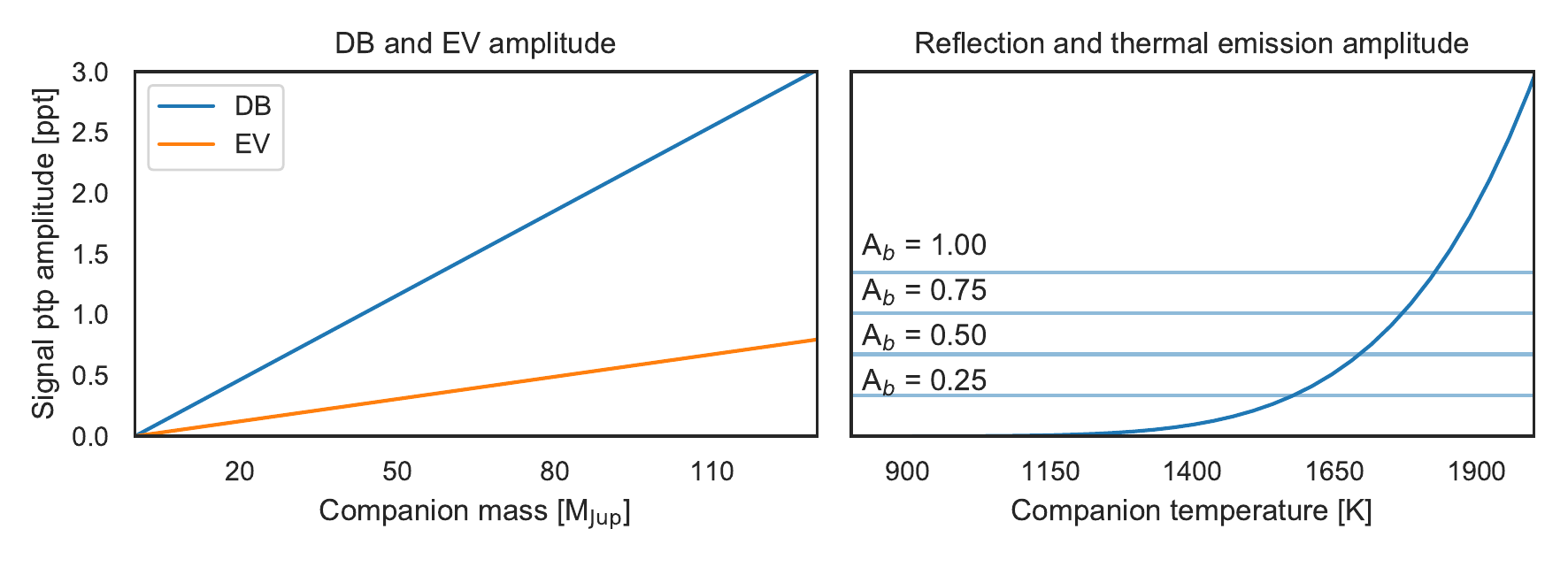}
		\caption{Amplitudes of the four phase curve components for \tplanet as a function of the unknown
		companion mass, Bond albedo ($A_\mathrm{b}$), and effective temperature.}
		\label{fig:phase_curve_amplitude}
	\end{figure*}

	The main four components contributing to the phase curve are Doppler boosting, ellipsoidal variations, 
	reflection, and thermal emission, as illustrated in Fig.~\ref{fig:phase_curve_schematic}. The relative importance
	of the different effects depends on the orbital geometry, the host star, and the companion properties. For example, the effects of Doppler
	boosting \citep{Loeb2003} are expected to be significantly more important than ellipsoidal 
	variations for \tplanet, as visible
	from Fig.~\ref{fig:phase_curve_amplitude} depicting the peak-to-peak expected amplitudes for \tplanet.

	Phase curve analysis can be useful in distinguishing between low-mass stars and substellar objects orbiting 
	low-mass stars on short-period orbits, and can also be used to distinguish between brown dwarfs and planets 
	if the photometric precision is sufficiently high. In our case, the host 
	star is faint by \tess standards, and we do not expect to detect significant phase variability if the companion
	is a planet or a low-mass brown dwarf. If the companion was a low-mass star, however, Doppler boosting
	should give rise to a clearly detectable signal. Our goal here is to derive upper limits for the amplitudes of 
	the different components, which can then be translated into upper limits for the effective temperature, 
	companion mass, and Bond albedo.
	
    We model the \tess out-of-transit light curve with a phase curve model combining reflection, thermal emission, Doppler
    boosting, and ellipsoidal variations. The \tess data are prepared the same way as for the main transit 
    light curve analysis, but we exclude the transits from the light curve. We also simplify the model
    slightly and fix the radius ratio, zero epoch, orbital period, orbital inclination, and semi-major axis 
    to their median posterior values derived by the multicolour transit analysis (the uncertainties left
    in these quantities after the multicolour modelling have a very minor effect on the phase curve model),
    and assume a circular orbit.
    
    The phase curve model is parameterised by the companion's Bond albedo, $A_\mathrm{B}$, log companion mass,
    $\log M_\mathrm{p}$, host star mass, $M_\star$, and the effective temperatures of the host and the 
    companion, $T_\star$ and $T_c$, respectively. In its most abstract form, the model is a sum of a
    constant baseline level $C$ ($\approx 1$) and the four components multiplied by their amplitudes
    \begin{equation}
         F(\phi) = C + A_\mathrm{r}F_\mathrm{r} + A_\mathrm{t}F_\mathrm{t} + A_\mathrm{b}F_\mathrm{b} + A_\mathrm{e}F_\mathrm{e},
    \end{equation}
    where $\phi \in [0,2\pi]$ is the orbital phase ($\phi=0$ for a transit), $A_\mathrm{r}F_\mathrm{r}$ is the 
    reflected light, $A_\mathrm{t}F_\mathrm{t}$ is thermal emission, $A_\mathrm{b}F_\mathrm{b}$ is the Doppler 
    boosting, and $A_\mathrm{e}F_\mathrm{e}$ is the ellipsoidal variation signal. 
    
    We approximate the planet as a Lambertian sphere\footnote{Lambertian reflectance is admittedly rather
    a poor reflectance model for a gas giant or a brown dwarf, but sufficient for our purposes here.} 
    \citep{Russell1916,Madhusudhan2011}, for which the phase function is given as 
    \begin{equation}
    A_\mathrm{r} F_\mathrm{r} = k^2 \frac{2}{3} \frac{A_\mathrm{B}}{a_\mathrm{s}^2} \times \frac{ (\sin\alpha + (\pi-\alpha) \cos\alpha)}{\pi}  \;\mathcal{E}(\phi),
    \end{equation}
    where $k$ is the planet-star radius ratio, $a_\mathrm{s}$ the scaled semi-major axis, $\alpha$ 
    the phase angle $\alpha = |\phi - \pi|$ , and $\mathcal{E}$ is the eclipse function that is
    modelled as a transit over an uniform disk but with depth scaled from 0 (full eclipse) to 1 (out of eclipse).

    The thermal emission is simplified to give a constant contribution to the observed flux over the whole orbital
    phase except when the companion is occulted by the star. The contribution is a product of planet-star area ratio
    and planet-star flux ratio calculated by approximating the host star and the companion as black bodies,
    \begin{equation}
       A_\mathrm{t}F_\mathrm{t} =  k^2 \frac{\int T(\lambda)\,P(T_\mathrm{p}, \lambda)\,\ud\lambda}{\int T(\lambda)\,P(T_\star, \lambda)\,\ud\lambda}\times \mathcal{E}(\phi),
    \end{equation}
    where $T$ is the \tess passband transmission, $P$ is Planck's law, and $\lambda$ the wavelength. 
    
    The expected Doppler boosting is calculated following \citet{Loeb2003}
    \begin{equation}
        A_\mathrm{b}F_\mathrm{b} = \frac{\beta}{c} \left( \frac{2\pi G}{p}\right)^{1/3} \frac{M_p \sin i}{M_\star^{2/3}} \times \sin\phi,
    \end{equation}
    where $c$ is the speed of light, $G$ is the gravitation constant, $p$ is the orbital period, and $\beta$ is the 
    photon-weighted passband-integrated beaming factor \citep{Bloemen2010}, described as 
     \begin{equation}
      \beta = \frac{\int T(\lambda)\,\lambda F_\lambda \,B\, \ud\lambda}{\int T(\lambda)\,\lambda F_\lambda\,\ud\lambda},
    \end{equation}
    where $F_\lambda$ is the stellar flux at wavelength $\lambda$, and $B = 5 + \ud \log F_\lambda / \ud \log \lambda$ 
    is the beaming factor \citep{Loeb2003}.\!\footnote{Methods to calculate the photon-weighted passband-integrated 
    beaming factor and the different phase curve components can be found from \pytransit.} The beaming
    factor is calculated based on a stellar spectrum modelled by \citet{Husser2013} rather than a black body 
    approximation.
    
    The ellipsoidal variation model follows from \citet{Pfahl2007} assuming a circular orbit, and is
    \begin{equation}
        A_\mathrm{e}F_\mathrm{e} = \frac{0.15 (15 + u) (1+g)}{3-g} \frac{M_p}{M_\star}  \left ( \frac{R_\star}{a}\right)^3 \times \sin^2 i \left(-\cos 2\phi\right),
    \end{equation}
    where $u$ is the linear limb darkening coefficient, $g$ is the gravity darkening coefficient, $a$ 
    is the semi-major axis, and $R_\star$ is the stellar radius.
    
    We model the correlated noise in the light curve as a Gaussian process (GP) following an 
    approximate Matern 3/2 kernel using the \texttt{celerite} package \citep{Foreman-Mackey2017}. This is because the 
    expected phase curve signal amplitudes are very small, and correlated noise could either lead to a false detection 
    or mask an existing real signal, and, especially, affect the posterior density tails. The flexibility from a GP model 
    leads to a conservative analysis where we can be secure we do not underestimate the component amplitudes allowed 
    by the data, and that the derived upper limits are robust. The GP noise model is parametrised by log white noise, 
    log input scale and log output scale, all of which are kept free in the optimisation and posterior estimation.
    
    We set a normal prior on the host star effective temperature of $\NP{3300, 100}$~K and uniform priors on the
    log companion mass (from 0.3~to~300$\,\mjup$, see also discussion about the posterior sensitivity on the prior and
    parametrisation in Sect.~\ref{sec:discussion}), effective temperature (from 500 to 3000~K), and Bond 
    albedo (from 0 to 1).  We set a normal prior on the log white noise standard deviation, $\NP{\hat{s}, 0.15}$,
    where $\hat{s}$ is a white noise estimate calculated from the flux point-to-point scatter. The log GP input scale
    has a uniform prior $\UP{-8,8}$, and the log output scale a wide normal prior $\NP{-6, 1.5}$.
    
	\section{Results}
	\label{sec:results}
	
    \begin{figure*}
		\centering
    	\includegraphics[width=\textwidth]{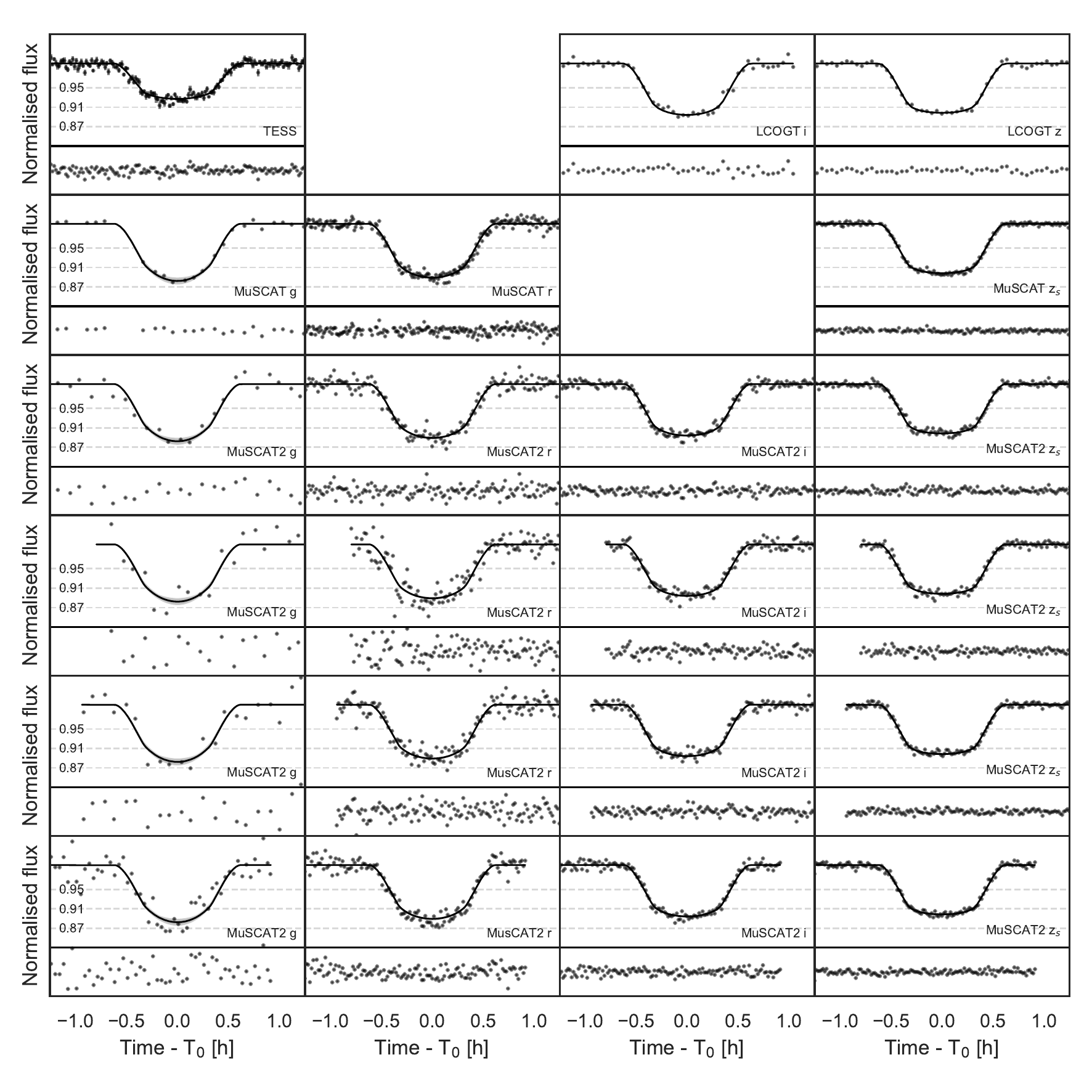}
		\caption{The phase folded and binned \tess light curve, MuSCAT2 light curves, MuSCAT light curves, 
		and LCOGT light curves together with the posterior median models and the residuals. The median baseline 
		model has been removed from the observed and modelled photometry for clarity.}
		\label{fig:light_curves}
	\end{figure*}
	
\begin{figure*}
		\centering
    	\includegraphics[width=\textwidth]{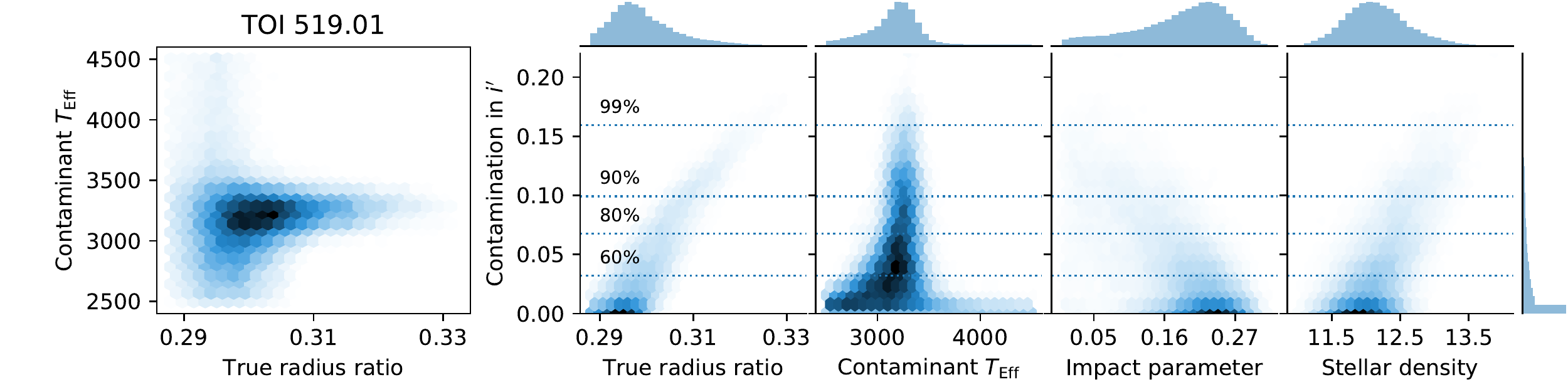}
		\caption{Marginal and joint posterior distributions for a set of parameters of interest from the 
		joint light curve analysis.}
		\label{fig:contamination_posteriors}
	\end{figure*}

\begin{figure}
		\centering
    	\includegraphics[width=\columnwidth]{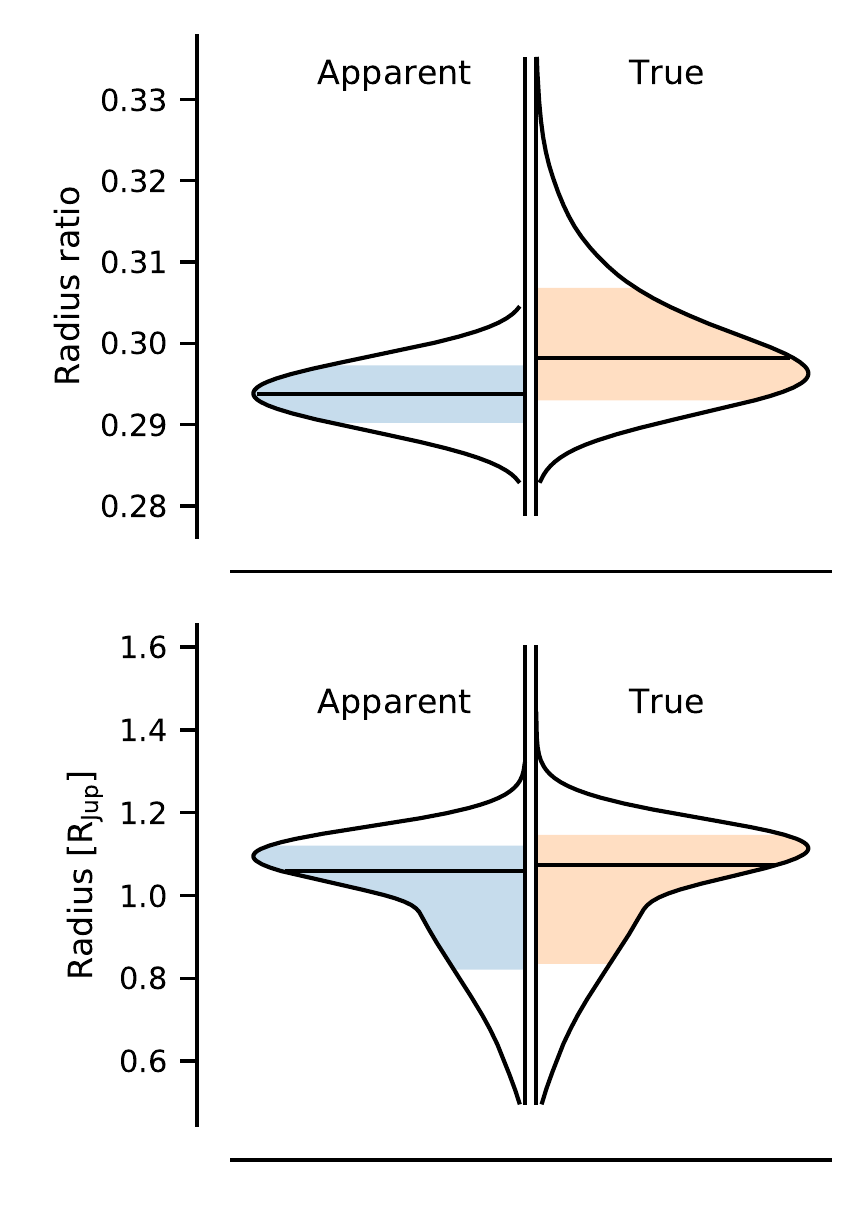}
		\caption{Marginal posterior distributions for the apparent and true radius ratios (top), and the
		apparent and true absolute planet radii (bottom). The true radius ratio posterior has a long tail
		towards large values due to possible flux contamination. However, its effect on the absolute radius
		estimate is minor due to the large uncertainties in the stellar radius estimate.}
		\label{fig:k_and_r_distributions}
	\end{figure}

	We show the photometry used in the multicolour analysis with the transit model in Fig.~\ref{fig:light_curves}, 
	the posterior densities
	for the true radius ratio, contaminant effective temperature, impact parameter, and stellar density
	in Fig.~\ref{fig:contamination_posteriors}, and the final marginal posterior 
	densities for the apparent and true radius ratio, and the apparent and true absolute radius in 
	Fig.~\ref{fig:k_and_r_distributions}.
    The multicolour analysis gives a \tess contamination estimate of $0.31_{-0.02}^{+0.04}$, allows us to exclude 
    significant flux contamination from sources of different spectral type than the host in the ground-based
    photometry, and also allows us to constrain the contamination from sources with a same spectral type as the host star to $< 15\%$
    in the ground-based photometry. 
    This leads to median true radius ratio of \ktmedian with a 5th percentile posterior lower limit of
    \ktlower and a 95th percentile posterior upper limit of \ktupper.

\begin{figure}
		\centering
    	\includegraphics[width=\columnwidth]{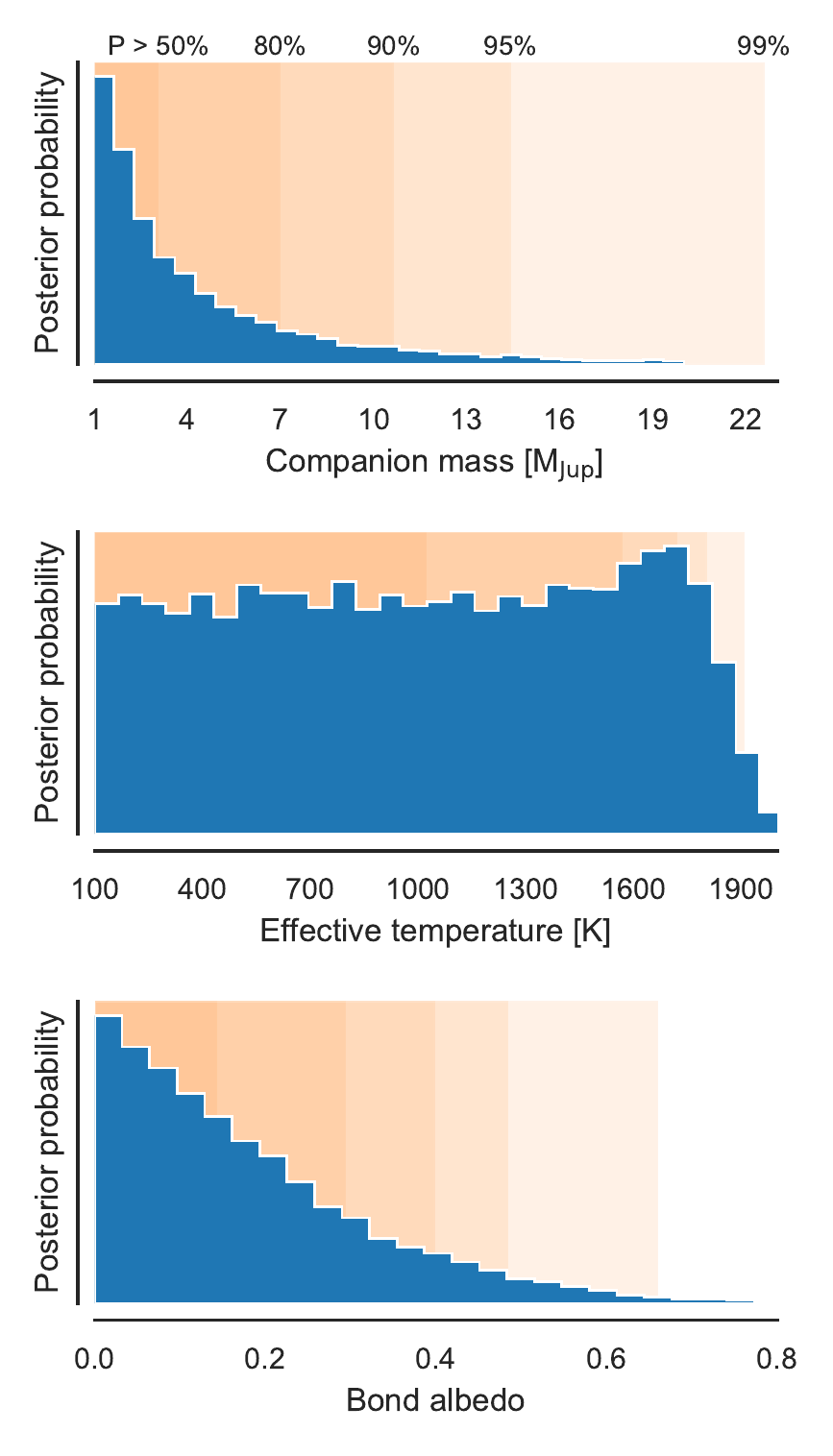}
		\caption{Marginal posterior distributions for the companion mass, effective temperature, and Bond 
		albedo. The orange shading shows a set of posterior percentile limits.}
		\label{fig:phase_model_posteriors}
	\end{figure}

   The posterior densities for the companion mass, effective temperature, and Bond albedo from the 
    phase curve analysis are shown in Fig.~\ref{fig:phase_model_posteriors}.
	The phase curve analysis leads to a \teffc posterior that is uniform
	between $0$~K and $1750$~K and then quickly slopes to zero. A tentative mode can be seen near 1700~K,
	but this is not statistically significant, and both the companion mass and Bond albedo have their modes
	at lower prior limit.  The analysis allows us to set an upper limit of 1800~K (corresponding to the
	95th posterior percentile) for the companion \teff, an upper albedo limit of 0.49, and upper companion mass 
	limit of 14\,\mjup. The companion mass posterior has a long tail, with a 99th percentile at $\sim22\,\mjup$.

	The companion mass posterior derived from the Doppler boosting and ellipsoidal variation signals can be sensitive on 
    the prior set on the mass. We parameterise the companion mass using log mass on which we set a uniform prior (which translates
    to a non-uniform prior on the mass), since the companion mass is a "scale" parameter with an unknown magnitude 
    \citep{Gelman2013,Parviainen2018}. We tested the posterior sensitivity on the parameterisation, and while the body of the
    posterior changes, the 95th posterior percentile is not affected significantly.
	
	\begin{table*}
		\centering
		\small
		\caption{Relative and absolute estimates for the stellar and companion parameters derived from the
		         multicolour transit analysis.}
		\begin{tabular*}{\textwidth}{@{\extracolsep{\fill}} llll}        
			\toprule\toprule
			\multicolumn{4}{l}{\emph{Ephemeris}} \\
			\midrule
			Transit epoch & $T_0$ & [BJD] & $ 2458491.8771169 \pm 1.3 \times 10^{-4}$\\
			Orbital period & $P$ & [days] &  $1.2652328 \pm  5\times 10^{-7}$ \\
			Transit duration & $T_{14}$ & [h] & $1.2332 \pm 0.0097$ \\
			\\
			\multicolumn{4}{l}{\emph{Relative properties}} \\
			\midrule
			Apparent radius ratio &$k_\mathrm{app}$ & $[R_\star]$ & $ 0.2939 \pm 0.0037$ \\
			True radius ratio &$k_\mathrm{true}$ & $[R_\star]$ & $0.30 \; (-0.0058) \; (+0.0091)$ \\
			Scaled semi-major axis &$a_\mathrm{s}$ & $[R_\star]$ & $10.11 \pm 0.14$ \\
			Impact parameter &$b$ && $0.19\;(-0.09)\;(+0.06)$ \\
			\\
			\multicolumn{4}{l}{\emph{Absolute properties}} \\
			\midrule 
			Apparent companion radius$^a$ &$R_{\mathrm{p,app}}$ & [\rjup]  &  $ 0.73 \pm 0.20 $ \\
			True companion radius$^a$ &$R_{\mathrm{p,true}}$ & [\rjup]  &  $ 0.75 \pm 0.21 $ \\
			Semi-major axis$^a$& $a$ &[AU] & $ 0.012 \pm 0.004 $\\
			Eq. temperature$^b$ & $T_{\mathrm{eq}}$ &[K] & $ 760 \pm 54 $ \\
			Stellar density & $\rho_\star$ & $[\gcm]$ & $12.20 \; (-0.44) \; (+0.55) $\\
			Inclination & $i$ &[deg] & $88.9 \pm 0.4$ \\
			\bottomrule       
		\end{tabular*}
		\tablefoot{ The estimates
			correspond to the posterior median ($P_{50}$) with $1 \sigma$ uncertainty estimate based on the 16th
			and 84th posterior percentiles ($P_{16}$ and $P_{84}$, respectively) for symmetric, 
			approximately normal posteriors. For asymmetric, unimodal posteriors, the estimates are $P_{50}{}^{P_{84}-P_{50}}_{P_{16}-P_{50}}$.
			\tablefoottext{a}{The semi-major axis and planet candidate radius are based on the scaled
				semi-major axis and true radius ratio samples, and the stellar radius estimate shown in Table~\ref{tbl:star}.} 
			\tablefoottext{b}{The equilibrium temperature of the planet candidate is calculated using the stellar \teff estimate, 
			scaled semi-major axis distribution, heat redistribution factor distributed uniformly between 0.25 and 0.5, and
			planetary albedo distributed uniformly between 0 and 0.4.}}
		\label{table:parameters}  
	\end{table*}
	
	\section{Discussion and Conclusions}
	\label{sec:discussion}
	
		\begin{figure*}
		\centering
    	\includegraphics[width=\textwidth]{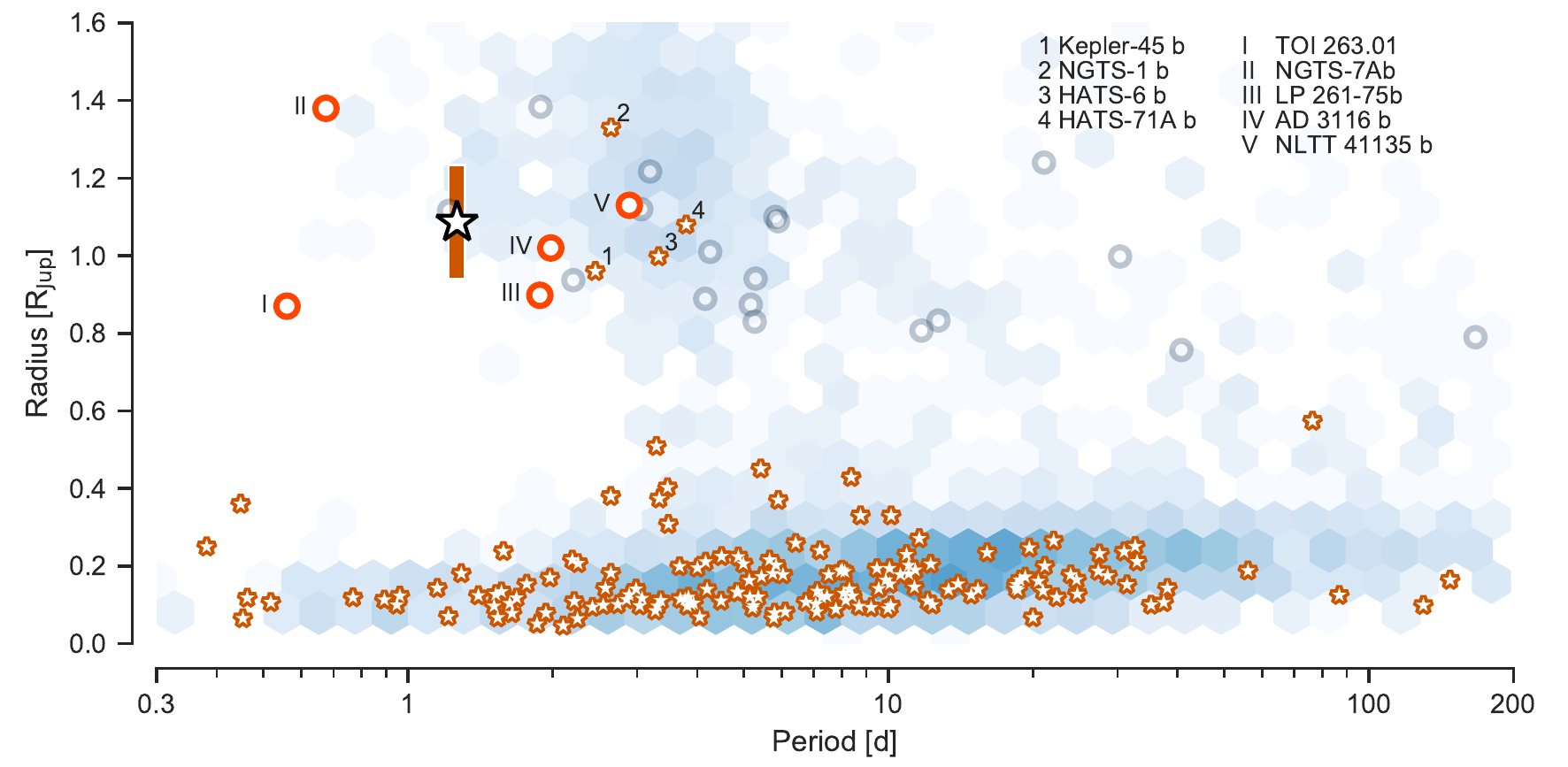}
    	\includegraphics[width=\textwidth]{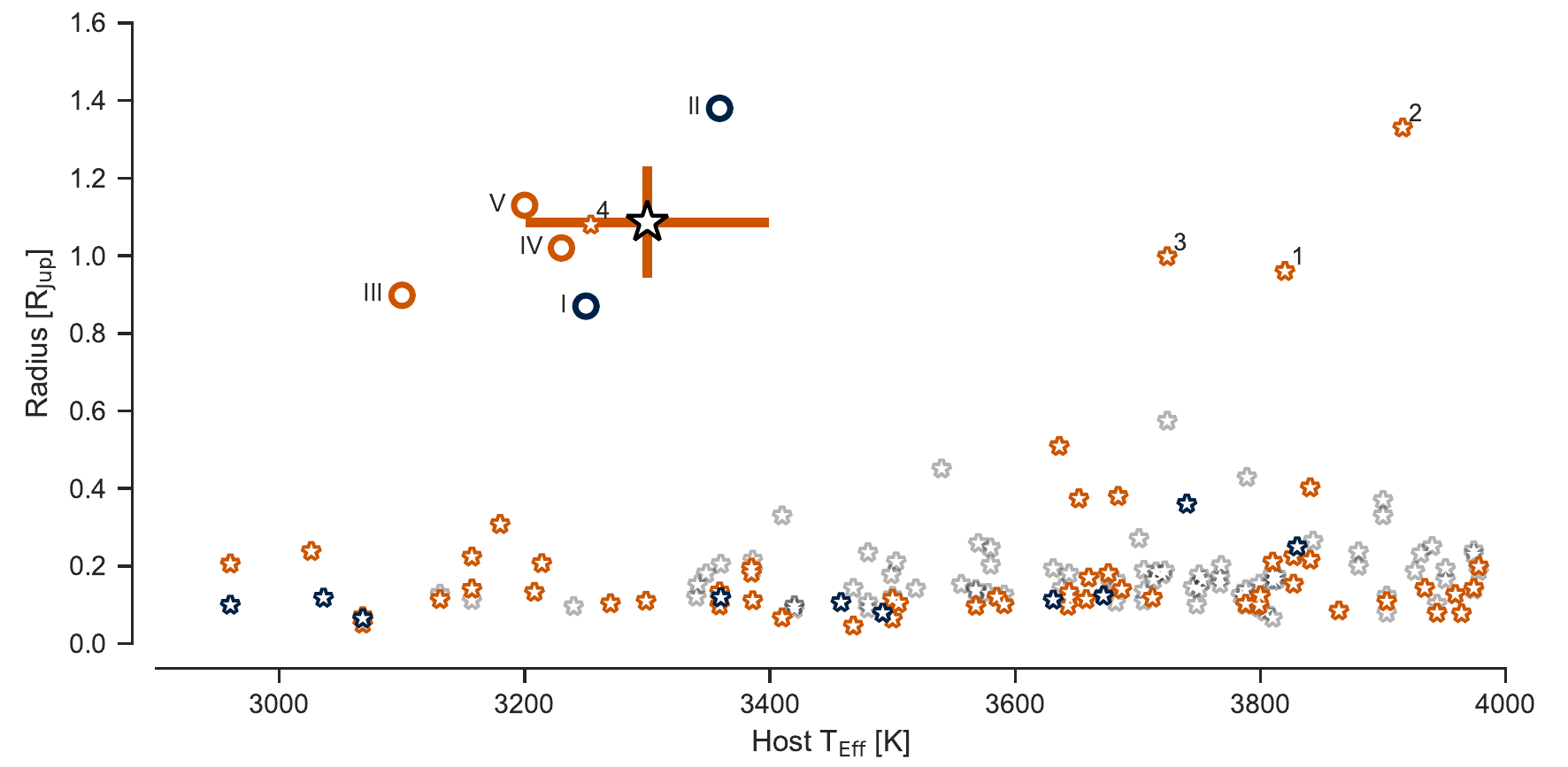}
		\caption{\tplanet in the context of currently known transiting planet and brown dwarf systems. First (top), we show the 
		radius as a function of orbital period for transiting planets and brown dwarfs (BDs) with a focus on companions around cool ($\teff<4000$~K) 
		host stars.  Transiting planets around cool hosts are shown as orange-rimmed stars, transiting BDs around cool hosts
		as orange-rimmed circles, transiting BDs around hot host ($\teff > 4000$~K) as dark-blue-rimmed circles, and transiting 
		planets around hot hosts as blue shading. Next (bottom), we show the radius as a function of the effective temperature 
		of the host star for transiting planets (stars) and brown dwarfs (circles). Objects with $P<1$~d are coloured as dark blue,
		$1<P<5$~d as orange, and $P>5$~d as light grey.}
		\label{fig:toi_519_in_context}
	\end{figure*}
    
	\begin{figure*}
		\centering
    	\includegraphics[width=\textwidth]{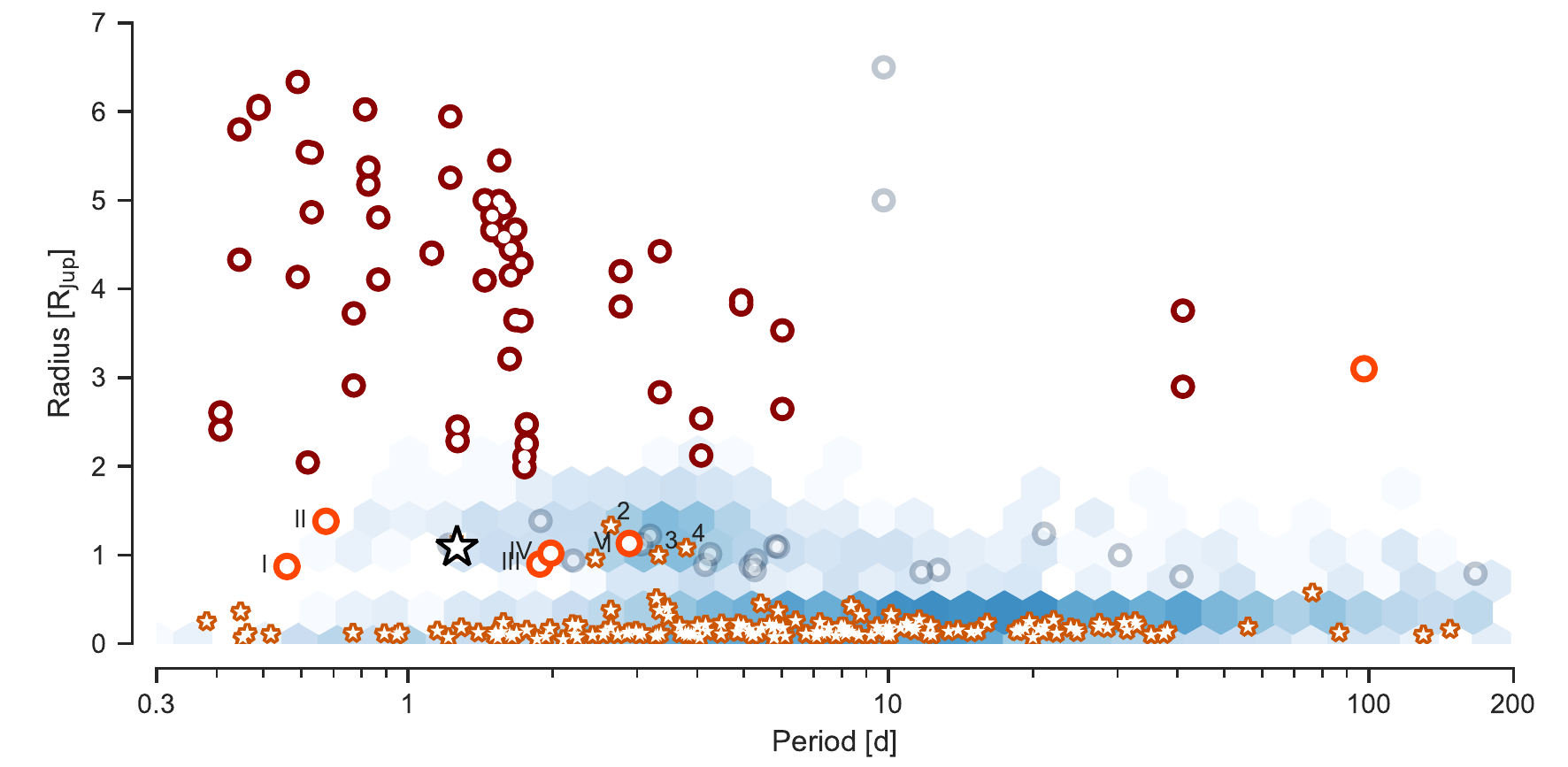}
		\caption{\tplanet in the context of currently known transiting planet and brown dwarf systems and eclipsing M~dwarf binaries
		marked as dark red circles. Otherwise as in Fig.~\ref{fig:toi_519_in_context}.}
		\label{fig:toi_519_in_context_with_m_binaries}
	\end{figure*}
	
	The reliability of transiting planet candidate validation based on constraining the size of the transiting object depends crucially on 
	the reliability of the stellar radius estimate. Large ($\sim 1 \rjup$) companions around low-mass stars are especially problematic due 
	to the mass-radius degeneracy for objects in this radius regime and uncertainties in the low-mass star radii. For \tplanet, while the 
	radius ratio is well-constrained, the companion's absolute radius depends on the stellar 
	radius estimate based on M~dwarf mass-radius relations and stellar classification based on low-resolution 
	spectroscopy template matching. The companion radius posterior ranges from  \rtlower~\rjup (certainly a planet) to 
	\rtupper~\rjup (low-mass star or young brown dwarf). Only with the effective temperature and companion mass limits from the phase
	curve analysis can we assess if the object is substellar.

	The radius, effective temperature, and mass constraints from the multicolour and phase curve analyses
	validate \tplanet as a substellar object. That is, with $R\sim1\,\rjup$, $M < 14\,\mjup$, and $T_mathrm{p} < 1800$~K, \tplanet 
	is either a very low mass brown dwarf or a massive planet located in a very sparsely populated region in the period-radius space for 
	substellar objects around M~dwarfs. Here, the mass limit from the Doppler boosting creates a stronger constrain on the nature
	of the companion than the temperature limit, since the system is likely old enough that even a low-mass star could have cooled down below our detection threshold. Further, the upper mass limit favours the hypothesis that \tplanet is a planet rather
	than a brown dwarf, but a brown dwarf scenario is strongly favoured a priori. The currently known hot Jupiters around M dwarfs generally orbit
	hosts hotter than 3700~K with periods larger than 2 days, with the exception of Kepler-45b (see Fig.~\ref{fig:toi_519_in_context}),
	and the (period, radius, host \teff) space \tplanet falls in is dominated by brown dwarfs.
	
    We caution that, in the case of \tstar, the \tess crowding metric appears to be overestimated by the 
	PDC pipeline. The \texttt{CROWDSAP} value that corresponds to the "ratio of target flux to the total flux in optimal aperture" 
	is 0.51, corresponding to a contamination of 0.49, while our derived \tess contamination posterior median is 0.31. Our \tess contamination
	estimate can be considered secure since it is directly related to the differences in the apparent transit depths measured from
	the \tess and ground-based light curves. The ground-based transits are shallower than expected based on the crowding-corrected 
	(PDC) \tess{}-observations, and this can only happen if the crowding is overestimated. Overestimated crowding leads to an overestimated
	radius ratio, and, thus, overestimated absolute radius. While this may be a rare occurrence, it is recommended to check for
	discrepancies between the \tess and ground-based photometry when carrying out a transit analysis, and include a free contamination
	factor for the \tess photometry if a joint transit analysis can be carried out with ground-based photometry.
	
	Figure~\ref{fig:toi_519_in_context} shows \tplanet in the context of all currently known planets and brown dwarfs transiting cool 
	host stars, and Fig.~\ref{fig:toi_519_in_context_with_m_binaries} extends these to include eclipsing M~dwarf binaries. While the 
	small number of objects does not allow to make any statistically significant inferences, we can recognise some curious features 
	whose significance will be uncovered in the future. First, all large ($R > 0.5\,\rjup$) objects orbiting cool ($\teff < 4000$~K) 
	dwarfs with periods shorter than 2~d are brown dwarfs. If \tplanet  is confirmed to be planet it will be the only inhabitant of this
	parameter space. However, this feature is not very significant in the current context of low-number statistics, since all four of
	the known hot Jupiters orbiting M~dwarfs have periods from 2 to 4~d (i.e also relatively short). More significant is the lack of
	large objects around cool dwarfs in the period range of 4 to 200~d, but this can be largely due to observational biases (the 
	figure includes only transiting objects with measured radii). 
    
    The lower panel of Figure~\ref{fig:toi_519_in_context} shows a radius versus host stellar temperature. We note that giant 
    planets and brown dwarfs seem to be found orbiting distinct spectral types. All large objects ($R > 0.5\,\rjup$) around the coolest
    dwarfs ($\teff<3400$~K) are brown dwarfs, except for HATS-71~A~b, and these  transiting brown dwarfs seem to be clustered
    orbiting cool dwarfs with $\teff\sim3100-3400$~K. On the contrary, giant planets are found around spectral types with $\teff>3700$~K. 
    For the spectral types with $\teff\sim3400-3700$~K there is a desert of any companions with a $R > 0.5\,\rjup$. Whether this apparent 
    clustering is of any significance needs to be verified by more hot Jupiter and brown dwarf discoveries around cool dwarfs.

	\begin{acknowledgements}
	We thank the anonymous referee for their helpful and constructive comments.
		We acknowledge financial support from the Agencia Estatal de Investigación of the Ministerio de Ciencia,
		Innovación y Universidades and the European FEDER/ERF funds through projects ESP2013-48391-C4-2-R, AYA2016-79425-C3-2-P, AYA2015-69350-C3-2-P, and PID2019-109522GB-C53, and PGC2018-098153-B-C31.
		This work is partly financed by the Spanish Ministry of Economics and Competitiveness through project ESP2016-80435-C2-2-R. We acknowledge supports by JSPS KAKENHI Grant Numbers JP17H04574, JP18H01265, and JP18H05439, and JST PRESTO Grant Number JPMJPR1775.
		JK acknowledges support by Deutsche Forschungsgemeinschaft (DFG) grants PA525/18-1 and PA525/19-1 within the DFG Schwerpunkt SPP 1992, Exploring the Diversity of Extra-solar Planets. MT is supported by JSPS KAKENHI grant Nos. 18H05442, 15H02063, and 22000005.
		This work was partly supported by Grant-in-Aid for JSPS Fellows, Grant Number JP20J21872.
		This article is partly based on observations made with the MuSCAT2 instrument, developed by ABC, at Telescopio Carlos S\'anchez operated on the island of Tenerife by the IAC in the Spanish Observatorio del Teide. We acknowledge the use of public \tess Alert data from pipelines at the \tess Science Office and at the \tess Science Processing Operations Center. This work makes use of observations from the LCOGT network. Resources supporting this work were provided by the NASA High-End Computing (HEC) Program through the NASA Advanced Supercomputing (NAS) Division at Ames Research Center for the production of the SPOC data products. This work makes use of observations from the LCOGT network.
	\end{acknowledgements}
    \clearpage

	\bibliographystyle{aa} 
	\bibliography{toi_519}

\begin{thebibliography}{84}
\expandafter\ifx\csname natexlab\endcsname\relax\def\natexlab#1{#1}\fi

\bibitem[{Almenara {et~al.}(2009)Almenara, Deeg, Aigrain, Alonso, Auvergne,
  Baglin, Barbieri, Barge, Bord{\'{e}}, Bouchy, Bruntt, Cabrera, Carone,
  Carpano, Catala, Csizmadia, {De la Reza}, Deleuil, Dvorak, Erikson, Fridlund,
  Gandolfi, Gillon, Gondoin, Guenther, Guillot, Hatzes, H{\'{e}}brard, Jorda,
  Lammer, L{\'{e}}ger, Llebaria, Loeillet, Magain, Mayor, Mazeh, Moutou,
  Ollivier, P{\"{a}}tzold, Pont, Queloz, Rauer, R{\'{e}}gulo, Renner, Rouan,
  Samuel, Schneider, Shporer, Wuchterl, \& Zucker}]{Almenara2009}
Almenara, J.~M., Deeg, H.~J., Aigrain, S., {et~al.} 2009, A{\&}A, 506, 337

\bibitem[{Angerhausen {et~al.}(2015)Angerhausen, DeLarme, \&
  Morse}]{Angerhausen2014}
Angerhausen, D., DeLarme, E., \& Morse, J.~A. 2015, Publ. Astron. Soc. Pacific,
  127, 1113

\bibitem[{Bakos {et~al.}(2018)Bakos, Bayliss, Bento, Bhatti, Brahm, Csubry,
  Espinoza, Hartman, Henning, Jord{\'{a}}n, Mancini, Penev, Rabus, Sarkis, Suc,
  de~Val-Borro, Zhou, Butler, Crane, Durkan, Shectman, Kim, L{\'{a}}z{\'{a}}r,
  Papp, S{\'{a}}ri, Ricker, Vanderspek, Latham, Seager, Winn, Jenkins, Chacon,
  Fűr{\'{e}}sz, Goeke, Li, Quinn, Quintana, Tenenbaum, Teske, Vezie, Yu,
  Stockdale, Evans, \& Relles}]{Bakos2018}
Bakos, G.~{\'{A}}., Bayliss, D., Bento, J., {et~al.} 2018
  [\eprint[arXiv]{1812.09406}]

\bibitem[{Barclay {et~al.}(2015)Barclay, Endl, Huber, Foreman-Mackey, Cochran,
  MacQueen, Rowe, \& Quintana}]{Barclay2014}
Barclay, T., Endl, M., Huber, D., {et~al.} 2015, Astrophys. J., 800, 46

\bibitem[{Barclay {et~al.}(2012)Barclay, Huber, Rowe, Fortney, Morley,
  Quintana, Fabrycky, Barentsen, Bloemen, Christiansen, Demory, Fulton,
  Jenkins, Mullally, Ragozzine, Seader, Shporer, Tenenbaum, \&
  Thompson}]{Barclay2012}
Barclay, T., Huber, D., Rowe, J.~F., {et~al.} 2012, Astrophys. J., 761, 53

\bibitem[{Bayliss {et~al.}(2018)Bayliss, Gillen, Eigm{\"{u}}ller, McCormac,
  Alexander, Armstrong, Booth, Bouchy, Burleigh, Cabrera, Casewell, Chaushev,
  Chazelas, Csizmadia, Erikson, Faedi, Foxell, G{\"{a}}nsicke, Goad, Grange,
  G{\"{u}}nther, Hodgkin, Jackman, Jenkins, Lambert, Louden, Metrailler,
  Moyano, Pollacco, Poppenhaeger, Queloz, Raddi, Rauer, Raynard, Smith, Soto,
  Thompson, Titz-Weider, Udry, Walker, Watson, West, \& Wheatley}]{Bayliss2018}
Bayliss, D., Gillen, E., Eigm{\"{u}}ller, P., {et~al.} 2018, MNRAS, 475, 4467

\bibitem[{Bloemen {et~al.}(2010)Bloemen, Marsh, {\O}stensen, Charpinet,
  Fontaine, Degroote, Heber, Kawaler, Aerts, Green, Telting, Brassard,
  G{\"{a}}nsicke, Handler, Kurtz, Silvotti, {Van Grootel}, Lindberg, Pursimo,
  Wilson, Gilliland, Kjeldsen, Christensen-Dalsgaard, Borucki, Koch, Jenkins,
  \& Klaus}]{Bloemen2010}
Bloemen, S., Marsh, T.~R., {\O}stensen, R.~H., {et~al.} 2010, MNRAS, 410, no

\bibitem[{Bradley {et~al.}(2019)Bradley, Sipocz, Robitaille, Tollerud,
  Vin{\'{i}}cius, Deil, Barbary, G{\"{u}}nther, Cara, Busko, Conseil,
  Droettboom, Bostroem, Bray, Bratholm, Wilson, Craig, Barentsen, Pascual,
  Donath, Greco, Perren, Lim, \& Kerzendorf}]{Bradley2019}
Bradley, L., Sipocz, B., Robitaille, T., {et~al.} 2019

\bibitem[{{Brown} {et~al.}(2013){Brown}, {Baliber}, {Bianco}, {Bowman},
  {Burleson}, {Conway}, {Crellin}, {Depagne}, {De Vera}, {Dilday}, {Dragomir},
  {Dubberley}, {Eastman}, {Elphick}, {Falarski}, {Foale}, {Ford}, {Fulton},
  {Garza}, {Gomez}, {Graham}, {Greene}, {Haldeman}, {Hawkins}, {Haworth},
  {Haynes}, {Hidas}, {Hjelstrom}, {Howell}, {Hygelund}, {Lister}, {Lobdill},
  {Martinez}, {Mullins}, {Norbury}, {Parrent}, {Paulson}, {Petry}, {Pickles},
  {Posner}, {Rosing}, {Ross}, {Sand}, {Saunders}, {Shobbrook}, {Shporer},
  {Street}, {Thomas}, {Tsapras}, {Tufts}, {Valenti}, {Vander Horst}, {Walker},
  {White}, \& {Willis}}]{Brown:2013}
{Brown}, T.~M., {Baliber}, N., {Bianco}, F.~B., {et~al.} 2013, Publications of
  the Astronomical Society of the Pacific, 125, 1031

\bibitem[{Burrows {et~al.}(2011)Burrows, Heng, \& Nampaisarn}]{Burrows2011}
Burrows, A.~S., Heng, K., \& Nampaisarn, T. 2011, Astrophys. J., 736, 47

\bibitem[{Cabrera {et~al.}(2017)Cabrera, Barros, Armstrong, Hidalgo, Santos,
  Almenara, Alonso, Deleuil, Demangeon, D{\'{i}}az, Lendl, Pfaff, Rauer,
  Santerne, Serrano, \& Zucker}]{Cabrera2017a}
Cabrera, J., Barros, S. C.~C., Armstrong, D., {et~al.} 2017, A{\&}A, 606, A75

\bibitem[{Cameron(2012)}]{Cameron2012}
Cameron, A.~C. 2012, Nature, 492, 48

\bibitem[{{Collins} {et~al.}(2017){Collins}, {Kielkopf}, {Stassun}, \&
  {Hessman}}]{Collins:2017}
{Collins}, K.~A., {Kielkopf}, J.~F., {Stassun}, K.~G., \& {Hessman}, F.~V.
  2017, \aj, 153, 77

\bibitem[{Dawson \& Johnson(2018)}]{Dawson2018}
Dawson, R.~I. \& Johnson, J.~A. 2018, Annu. Rev. Astron. Astrophys., 56, 175

\bibitem[{Dressing \& Charbonneau(2015)}]{Dressing2015}
Dressing, C.~D. \& Charbonneau, D. 2015, Astrophys. J., 807, 45

\bibitem[{Esteves {et~al.}(2013)Esteves, {De Mooij}, \&
  Jayawardhana}]{Esteves2013}
Esteves, L.~J., {De Mooij}, E. J.~W., \& Jayawardhana, R. 2013, Astrophys. J.,
  772, 51

\bibitem[{Esteves {et~al.}(2015)Esteves, Mooij, \& Jayawardhana}]{Esteves2015}
Esteves, L.~J., Mooij, E. J. W.~D., \& Jayawardhana, R. 2015, Astrophys. J.,
  804, 150

\bibitem[{Feinstein {et~al.}(2019)Feinstein, Montet, Foreman-Mackey, Bedell,
  Saunders, Bean, Christiansen, Hedges, Luger, Scolnic, \&
  Cardoso}]{Feinstein2019}
Feinstein, A.~D., Montet, B.~T., Foreman-Mackey, D., {et~al.} 2019, Publ.
  Astron. Soc. Pacific, 131 [\eprint[arXiv]{1903.09152}]

\bibitem[{Foreman-Mackey {et~al.}(2017)Foreman-Mackey, Agol, Ambikasaran, \&
  Angus}]{Foreman-Mackey2017}
Foreman-Mackey, D., Agol, E., Ambikasaran, S., \& Angus, R. 2017, Astron. J.,
  154, 220

\bibitem[{Foreman-Mackey {et~al.}(2013)Foreman-Mackey, Hogg, Lang, \&
  Goodman}]{Foreman-Mackey2012}
Foreman-Mackey, D., Hogg, D.~W., Lang, D., \& Goodman, J. 2013, Publ. Astron.
  Soc. Pacific, 125, 306

\bibitem[{Fressin {et~al.}(2013)Fressin, Torres, Charbonneau, Bryson,
  Christiansen, Dressing, Jenkins, Walkowicz, \& Batalha}]{Fressin2013}
Fressin, F., Torres, G., Charbonneau, D., {et~al.} 2013, Astrophys. J., 766, 81

\bibitem[{Fukui {et~al.}(2011)Fukui, Narita, Tristram, Sumi, Abe, Itow,
  Sullivan, Bond, Hirano, Tamura, Bennett, Furusawa, Hayashi, Hearnshaw,
  Hosaka, Kamiya, Kobara, Korpela, Kilmartin, Lin, Ling, Makita, Masuda,
  Matsubara, Miyake, Muraki, Nagaya, Nishimoto, Ohnishi, Omori, Perrott,
  Rattenbury, Saito, Skuljan, Suzuki, Sweatman, \& Wada}]{Fukui2011}
Fukui, A., Narita, N., Tristram, P.~J., {et~al.} 2011, Publ. Astron. Soc.
  Japan, 63, 287

\bibitem[{Gelman {et~al.}(2013)Gelman, Carlin, Stern, Dunson, Vehtari, \&
  Rubin}]{Gelman2013}
Gelman, A., Carlin, J.~B., Stern, H.~S., {et~al.} 2013, {Bayesian data
  analysis, third edition}, 3rd edn. (CRC Press), 1--646

\bibitem[{Gillen {et~al.}(2017)Gillen, Hillenbrand, David, Aigrain, Rebull,
  Stauffer, Cody, \& Queloz}]{Gillen2017}
Gillen, E., Hillenbrand, L.~A., David, T.~J., {et~al.} 2017, Astrophys. J.,
  849, 11

\bibitem[{{Gizis}(1997)}]{gizis97}
{Gizis}, J.~E. 1997, \aj, 113, 806

\bibitem[{Goodman \& Weare(2010)}]{Goodman2010}
Goodman, J. \& Weare, J. 2010, Commun. Appl. Math. Comput. Sci., 5, 65

\bibitem[{Hartman {et~al.}(2015)Hartman, Bayliss, Brahm, Bakos, Mancini,
  Jord{\`{a}}n, Penev, Rabus, Zhou, Butler, Espinoza, {De Val-Borro}, Bhatti,
  Csubry, Ciceri, Henning, Schmidt, Arriagada, Shectman, Crane, Thompson, Suc,
  Cs{\`{a}}k, Tan, Noyes, L{\`{a}}z{\`{a}}r, Papp, \& S{\`{a}}ri}]{Hartman2015}
Hartman, J.~D., Bayliss, D., Brahm, R., {et~al.} 2015, Astron. J., 149, 166

\bibitem[{{Houdebine} {et~al.}(2019){Houdebine}, {Mullan}, {Doyle}, {de La
  Vieuville}, {Butler}, \& {Paletou}}]{houdebine19}
{Houdebine}, {\'E}.~R., {Mullan}, D.~J., {Doyle}, J.~G., {et~al.} 2019, \aj,
  158, 56

\bibitem[{Hoyer \& Hamman(2017)}]{Hoyer2017}
Hoyer, S. \& Hamman, J.~J. 2017, J. Open Res. Softw., 5, 1

\bibitem[{Hunter(2007)}]{Hunter2007}
Hunter, J.~D. 2007, Comput. Sci. Eng., 9, 90

\bibitem[{Husser {et~al.}(2013)Husser, {Wende-von Berg}, Dreizler, Homeier,
  Reiners, Barman, \& Hauschildt}]{Husser2013}
Husser, T.-O., {Wende-von Berg}, S., Dreizler, S., {et~al.} 2013, A{\&}A, 553,
  A6

\bibitem[{Irwin {et~al.}(2010)Irwin, Buchhave, Berta, Charbonneau, Latham,
  Burke, Esquerdo, Everett, Holman, Nutzman, Berlind, Calkins, Falco, Winn,
  Johnson, \& Gazak}]{Irwin2010}
Irwin, J., Buchhave, L., Berta, Z.~K., {et~al.} 2010, Astrophys. J., 718, 1353

\bibitem[{Irwin {et~al.}(2018)Irwin, Charbonneau, Esquerdo, Latham, Winters,
  Dittmann, Newton, Berta-Thompson, Berlind, \& Calkins}]{Irwin2018}
Irwin, J.~M., Charbonneau, D., Esquerdo, G.~A., {et~al.} 2018, Astron. J., 156,
  140

\bibitem[{Jackman {et~al.}(2019)Jackman, Wheatley, Bayliss, Gill, Hodgkin,
  Burleigh, Braker, G{\"{u}}nther, Louden, Turner, Anderson, Belardi, Bouchy,
  Briegal, Bryant, Cabrera, Casewell, Chaushev, Costes, Csizmadia,
  Eigm{\"{u}}ller, Erikson, G{\"{a}}nsicke, Gillen, Goad, Jenkins, McCormac,
  Moyano, Nielsen, Pollacco, Poppenhaeger, Queloz, Rauer, Raynard, Smith, Udry,
  Vines, Watson, \& West}]{Jackman2019}
Jackman, J. A.~G., Wheatley, P.~J., Bayliss, D., {et~al.} 2019, MNRAS, 489,
  5146

\bibitem[{Jenkins {et~al.}(2016)Jenkins, Twicken, McCauliff, Campbell,
  Sanderfer, Lung, Mansouri-Samani, Girouard, Tenenbaum, Klaus, Smith,
  Caldwell, Chacon, Henze, Heiges, Latham, Morgan, Swade, Rinehart, \&
  Vanderspek}]{Jenkins2016}
Jenkins, J.~M., Twicken, J.~D., McCauliff, S., {et~al.} 2016, in Proc. SPIE,
  ed. G.~Chiozzi \& J.~C. Guzman, 99133E

\bibitem[{{Jensen}(2013)}]{Jensen:2013}
{Jensen}, E. 2013, {Tapir: A web interface for transit/eclipse observability},
  Astrophysics Source Code Library

\bibitem[{Johnson {et~al.}(2012)Johnson, Gazak, Apps, Muirhead, Crepp,
  Crossfield, {Tabetha Boyajian}, {Von Braun}, Rojas-Ayala, Howard, Covey,
  Schlawin, Hamren, Morton, Marcy, \& Lloyd}]{Johnson2012}
Johnson, J.~A., Gazak, J.~Z., Apps, K., {et~al.} 2012, Astron. J., 143
  [\eprint[arXiv]{1112.0017}]

\bibitem[{Kesseli {et~al.}(2017)Kesseli, West, Veyette, Harrison, Feldman, \&
  Bochanski}]{Kesseli2017}
Kesseli, A.~Y., West, A.~A., Veyette, M., {et~al.} 2017, Astrophys. J. Suppl.
  Ser., 230, 16

\bibitem[{{Kirkpatrick} {et~al.}(1991){Kirkpatrick}, {Henry}, \&
  {McCarthy}}]{kirk91}
{Kirkpatrick}, J.~D., {Henry}, T.~J., \& {McCarthy}, Donald~W., J. 1991, \apjs,
  77, 417

\bibitem[{Kirkpatrick {et~al.}(2014)Kirkpatrick, Schneider, Fajardo-Acosta,
  Gelino, Mace, Wright, Logsdon, McLean, Cushing, Skrutskie, Eisenhardt, Stern,
  Balokovi{\'{c}}, Burgasser, Faherty, Lansbury, Rich, Skrzypek, Fowler, Cutri,
  Masci, Conrow, Grillmair, McCallon, Beichman, \& Marsh}]{Kirkpatrick2014}
Kirkpatrick, J.~D., Schneider, A., Fajardo-Acosta, S., {et~al.} 2014,
  Astrophys. J., 783

\bibitem[{Lang {et~al.}(2010)Lang, Hogg, Mierle, Blanton, \& Roweis}]{Lang2010}
Lang, D., Hogg, D.~W., Mierle, K., Blanton, M., \& Roweis, S. 2010, Astron. J.,
  139, 1782

\bibitem[{Lillo-Box {et~al.}(2014)Lillo-Box, Barrado, Moya, Montesinos,
  Montalb{\'{a}}n, Bayo, Barbieri, R{\'{e}}gulo, Mancini, Bouy, \&
  Henning}]{Lillo-Box2014}
Lillo-Box, J., Barrado, D., Moya, A., {et~al.} 2014, A{\&}A, 562, A109

\bibitem[{Loeb \& Gaudi(2003)}]{Loeb2003}
Loeb, A. \& Gaudi, B.~S. 2003, Astrophys. J., 588, L117

\bibitem[{Madhusudhan \& Burrows(2012)}]{Madhusudhan2011}
Madhusudhan, N. \& Burrows, A.~S. 2012, {ANALYTIC MODELS FOR ALBEDOS, PHASE
  CURVES, AND POLARIZATION OF REFLECTED LIGHT FROM EXOPLANETS}

\bibitem[{Mann {et~al.}(2019)Mann, Dupuy, Kraus, Gaidos, Ansdell, Ireland,
  Rizzuto, Hung, Dittmann, Factor, Feiden, Martinez,
  Ru{\'{i}}z-Rodr{\'{i}}guez, \& {Chia Thao}}]{Mann2019}
Mann, A.~W., Dupuy, T., Kraus, A.~L., {et~al.} 2019, Astrophys. J., 871, 63

\bibitem[{{Martin} {et~al.}(1996){Martin}, {Rebolo}, \&
  {Zapatero-Osorio}}]{martin96}
{Martin}, E.~L., {Rebolo}, R., \& {Zapatero-Osorio}, M.~R. 1996, \apj, 469, 706

\bibitem[{Mazeh \& Faigler(2010)}]{Mazeh2010}
Mazeh, T. \& Faigler, S. 2010, A{\&}A, 521, L59

\bibitem[{{McCully} {et~al.}(2018){McCully}, {Volgenau}, {Harbeck}, {Lister},
  {Saunders}, {Turner}, {Siiverd}, \& {Bowman}}]{McCully:2018}
{McCully}, C., {Volgenau}, N.~H., {Harbeck}, D.-R., {et~al.} 2018, in Society
  of Photo-Optical Instrumentation Engineers (SPIE) Conference Series, Vol.
  10707, \procspie, 107070K

\bibitem[{Mckinney(2010)}]{Mckinney2010}
Mckinney, W. 2010in , 51--56

\bibitem[{Mislis {et~al.}(2012)Mislis, Heller, Schmitt, \&
  Hodgkin}]{Mislis2012a}
Mislis, D., Heller, R., Schmitt, J. H. M.~M., \& Hodgkin, S. 2012, A{\&}A, 538,
  A4

\bibitem[{Mislis \& Hodgkin(2012)}]{Mislis2012}
Mislis, D. \& Hodgkin, S. 2012, MNRAS, 422, 1512

\bibitem[{Mordasini {et~al.}(2012)Mordasini, Alibert, Benz, Klahr, \&
  Henning}]{Mordasini2012}
Mordasini, C., Alibert, Y., Benz, W., Klahr, H., \& Henning, T. 2012, A{\&}A,
  541, A97

\bibitem[{Moutou {et~al.}(2009)Moutou, Pont, Bouchy, Deleuil, Almenara, Alonso,
  Barbieri, Bruntt, Deeg, Fridlund, Gandolfi, Gillon, Guenther, Hatzes,
  H{\'{e}}brard, Loeillet, Mayor, Mazeh, Queloz, Rabus, Rouan, Shporer, Udry,
  Aigrain, Auvergne, Baglin, Barge, Benz, Bord{\'{e}}, Carpano, {De la Reza},
  Dvorak, Erikson, Gondoin, Guillot, Jorda, Kabath, Lammer, L{\'{e}}ger,
  Llebaria, Lovis, Magain, Ollivier, P{\"{a}}tzold, Pepe, Rauer, Schneider, \&
  Wuchterl}]{Moutou2009}
Moutou, C., Pont, F., Bouchy, F., {et~al.} 2009, A{\&}A, 506, 321

\bibitem[{Mullally {et~al.}(2018)Mullally, Thompson, Coughlin, Burke, \&
  Rowe}]{Mullally2018}
Mullally, F., Thompson, S.~E., Coughlin, J.~L., Burke, C.~J., \& Rowe, J.~F.
  2018, Astron. J., 155, 210

\bibitem[{Narita {et~al.}(2015)Narita, Fukui, Kusakabe, Onitsuka, Ryu,
  Yanagisawa, Izumiura, Tamura, \& Yamamuro}]{Narita2015}
Narita, N., Fukui, A., Kusakabe, N., {et~al.} 2015, J. Astron. Telesc.
  Instruments, Syst., 1, 045001

\bibitem[{Narita {et~al.}(2019)Narita, Fukui, Kusakabe, Watanabe, Palle,
  Parviainen, Monta{\~{n}}{\'{e}}s-Rodr{\'{i}}guez, Murgas, Monelli, Aguiar, \&
  {Perez Prieto}}]{Narita2018}
Narita, N., Fukui, A., Kusakabe, N., {et~al.} 2019, J. Astron. Telesc.
  Instruments, Syst., 5, 015001

\bibitem[{Parviainen(2015)}]{Parviainen2015}
Parviainen, H. 2015, MNRAS, 450, 3233

\bibitem[{Parviainen(2018)}]{Parviainen2018}
Parviainen, H. 2018, in Handb. Exopl. (Cham: Springer International
  Publishing), 1--24

\bibitem[{Parviainen \& Aigrain(2015)}]{Parviainen2015b}
Parviainen, H. \& Aigrain, S. 2015, MNRAS, 453, 3822

\bibitem[{Parviainen {et~al.}(2020)Parviainen, Palle, Zapatero-Osorio,
  {Montanes Rodriguez}, Murgas, Narita, {Hidalgo Soto}, B{\'{e}}jar, Korth,
  Monelli, {Casasayas Barris}, Crouzet, de~Leon, Fukui, Hernandez, Klagyivik,
  Kusakabe, Luque, Mori, Nishiumi, Prieto-Arranz, Tamura, Watanabe, Burke,
  Charbonneau, Collins, Collins, Conti, {Garcia Soto}, Jenkins, Jenkins,
  Levine, Li, Rinehart, Seager, Tenenbaum, Ting, Vanderspek, Vezie, \&
  Winn}]{Parviainen2020}
Parviainen, H., Palle, E., Zapatero-Osorio, M.~R., {et~al.} 2020, A{\&}A, 633,
  A28

\bibitem[{Parviainen {et~al.}(2019)Parviainen, Tingley, Deeg, Palle, Alonso,
  {Montanes Rodriguez}, Murgas, Narita, Fukui, Watanabe, Kusakabe, Tamura,
  Nishiumi, Prieto-Arranz, Klagyivik, B{\'{e}}jar, Crouzet, Mori, {Hidalgo
  Soto}, {Casasayas Barris}, \& Luque}]{Parviainen2019}
Parviainen, H., Tingley, B., Deeg, H.~J., {et~al.} 2019, A{\&}A, 630, A89

\bibitem[{Perez \& Granger(2007)}]{Perez2007}
Perez, F. \& Granger, B. 2007, Comput. Sci. Eng., 21

\bibitem[{Pfahl {et~al.}(2008)Pfahl, Arras, \& Paxton}]{Pfahl2007}
Pfahl, E., Arras, P., \& Paxton, B. 2008, Astrophys. J., 679, 783

\bibitem[{Price {et~al.}(2005)Price, Storn, \& Lampinen}]{Price2005}
Price, K., Storn, R., \& Lampinen, J. 2005, {Differential Evolution} (Berlin:
  Springer)

\bibitem[{Price-Whelan {et~al.}(2018)Price-Whelan, Sipőcz, G{\"{u}}nther, Lim,
  Crawford, Conseil, Shupe, Craig, Dencheva, Ginsburg, VanderPlas, Bradley,
  P{\'{e}}rez-Su{\'{a}}rez, de~Val-Borro, Aldcroft, Cruz, Robitaille, Tollerud,
  Ardelean, Babej, Bach, Bachetti, Bakanov, Bamford, Barentsen, Barmby,
  Baumbach, Berry, Biscani, Boquien, Bostroem, Bouma, Brammer, Bray,
  Breytenbach, Buddelmeijer, Burke, Calderone, Rodr{\'{i}}guez, Cara, Cardoso,
  Cheedella, Copin, Corrales, Crichton, D'Avella, Deil, Depagne, Dietrich,
  Donath, Droettboom, Earl, Erben, Fabbro, Ferreira, Finethy, Fox, Garrison,
  Gibbons, Goldstein, Gommers, Greco, Greenfield, Groener, Grollier, Hagen,
  Hirst, Homeier, Horton, Hosseinzadeh, Hu, Hunkeler, Ivezi{\'{c}}, Jain,
  Jenness, Kanarek, Kendrew, Kern, Kerzendorf, Khvalko, King, Kirkby, Kulkarni,
  Kumar, Lee, Lenz, Littlefair, Ma, Macleod, Mastropietro, McCully, Montagnac,
  Morris, Mueller, Mumford, Muna, Murphy, Nelson, Nguyen, Ninan, N{\"{o}}the,
  Ogaz, Oh, Parejko, Parley, Pascual, Patil, Patil, Plunkett, Prochaska,
  Rastogi, Janga, Sabater, Sakurikar, Seifert, Sherbert, Sherwood-Taylor, Shih,
  Sick, Silbiger, Singanamalla, Singer, Sladen, Sooley, Sornarajah, Streicher,
  Teuben, Thomas, Tremblay, Turner, Terr{\'{o}}n, van Kerkwijk, de~la Vega,
  Watkins, Weaver, Whitmore, Woillez, \& Zabalza}]{Price-Whelan2018}
Price-Whelan, A.~M., Sipőcz, B.~M., G{\"{u}}nther, H.~M., {et~al.} 2018,
  Astron. J., 156, 123

\bibitem[{Quintana {et~al.}(2013)Quintana, Rowe, Barclay, Howell, Ciardi,
  Demory, Caldwell, Borucki, Christiansen, Jenkins, Klaus, Fulton, Morris,
  Sanderfer, Shporer, Smith, Still, \& Thompson}]{Quintana2013}
Quintana, E.~V., Rowe, J.~F., Barclay, T., {et~al.} 2013, Astrophys. J., 767,
  137

\bibitem[{Ricker {et~al.}(2014)Ricker, Winn, Vanderspek, Latham, Bakos, Bean,
  Berta-Thompson, Brown, Buchhave, Butler, Butler, Chaplin, Charbonneau,
  Christensen-Dalsgaard, Clampin, Deming, Doty, {De Lee}, Dressing, Dunham,
  Endl, Fressin, Ge, Henning, Holman, Howard, Ida, Jenkins, Jernigan, Johnson,
  Kaltenegger, Kawai, Kjeldsen, Laughlin, Levine, Lin, Lissauer, MacQueen,
  Marcy, McCullough, Morton, Narita, Paegert, Palle, Pepe, Pepper, Quirrenbach,
  Rinehart, Sasselov, Sato, Seager, Sozzetti, Stassun, Sullivan, Szentgyorgyi,
  Torres, Udry, \& Villasenor}]{Ricker2014}
Ricker, G.~R., Winn, J.~N., Vanderspek, R., {et~al.} 2014, J. Astron. Telesc.
  Instruments, Syst., 1, 014003

\bibitem[{Russell(1916)}]{Russell1916}
Russell, H.~N. 1916, Astrophys. J., 43, 173

\bibitem[{Santerne {et~al.}(2012)Santerne, D{\'{i}}az, Moutou, Bouchy,
  H{\'{e}}brard, Almenara, Bonomo, Deleuil, \& Santos}]{Santerne2012}
Santerne, A., D{\'{i}}az, R.~F., Moutou, C., {et~al.} 2012, A{\&}A, 545, A76

\bibitem[{Schweitzer {et~al.}(2019)Schweitzer, Passegger, Cifuentes,
  B{\'{e}}jar, Cort{\'{e}}s-Contreras, Caballero, del Burgo, Czesla,
  K{\"{u}}rster, Montes, {Zapatero Osorio}, Ribas, Reiners, Quirrenbach, Amado,
  Aceituno, Anglada-Escud{\'{e}}, Bauer, Dreizler, Jeffers, Guenther, Henning,
  Kaminski, Lafarga, Marfil, Morales, Schmitt, Seifert, Solano, Tabernero, \&
  Zechmeister}]{Schweitzer2019}
Schweitzer, A., Passegger, V.~M., Cifuentes, C., {et~al.} 2019, A{\&}A, 625,
  A68

\bibitem[{Shporer(2017)}]{Shporer2017}
Shporer, A. 2017, Publ. Astron. Soc. Pacific, 129, 72001

\bibitem[{Shporer {et~al.}(2011)Shporer, Jenkins, Rowe, Sanderfer, Seader,
  Smith, Still, Thompson, Twicken, \& Welsh}]{Shporer2011a}
Shporer, A., Jenkins, J.~M., Rowe, J.~F., {et~al.} 2011, Astron. J., 142, 195

\bibitem[{Shporer {et~al.}(2019)Shporer, Wong, Huang, Line, Stassun, Fetherolf,
  Kane, Bouma, Daylan, G{\"{u}}enther, Ricker, Latham, Vanderspek, Seager,
  Winn, Jenkins, Glidden, Berta-Thompson, Ting, Li, \& Haworth}]{Shporer2019}
Shporer, A., Wong, I., Huang, C.~X., {et~al.} 2019, Astron. J., 157, 178

\bibitem[{Smith {et~al.}(2012)Smith, Stumpe, {Van Cleve}, Jenkins, Barclay,
  Fanelli, Girouard, Kolodziejczak, McCauliff, Morris, \& Twicken}]{Smith2012a}
Smith, J.~C., Stumpe, M.~C., {Van Cleve}, J.~E., {et~al.} 2012, Publ. Astron.
  Soc. Pacific, 124, 1000

\bibitem[{Stassun {et~al.}(2017{\natexlab{a}})Stassun, Collins, \&
  Gaudi}]{Stassun2017a}
Stassun, K.~G., Collins, K.~A., \& Gaudi, B.~S. 2017{\natexlab{a}}, Astron. J.,
  153, 136

\bibitem[{Stassun {et~al.}(2017{\natexlab{b}})Stassun, Corsaro, Pepper, \&
  Gaudi}]{Stassun2017}
Stassun, K.~G., Corsaro, E., Pepper, J.~A., \& Gaudi, B.~S. 2017{\natexlab{b}},
  Astron. J., 155, 22

\bibitem[{Stassun \& Torres(2016)}]{Stassun2016}
Stassun, K.~G. \& Torres, G. 2016, Astron. J., 152, 180

\bibitem[{Stassun \& Torres(2018)}]{Stassun2018}
Stassun, K.~G. \& Torres, G. 2018, Astrophys. J., 862, 61

\bibitem[{Storn \& Price(1997)}]{Storn1997}
Storn, R. \& Price, K. 1997, J. Glob. Optim., 11, 341

\bibitem[{Stumpe {et~al.}(2014)Stumpe, Smith, Catanzarite, {Van Cleve},
  Jenkins, Twicken, \& Girouard}]{Stumpe2014}
Stumpe, M.~C., Smith, J.~C., Catanzarite, J.~H., {et~al.} 2014, Publ. Astron.
  Soc. Pacific, 126, 100

\bibitem[{Stumpe {et~al.}(2012)Stumpe, Smith, {Van Cleve}, Twicken, Barclay,
  Fanelli, Girouard, Jenkins, Kolodziejczak, McCauliff, \& Morris}]{Stumpe2012}
Stumpe, M.~C., Smith, J.~C., {Van Cleve}, J.~E., {et~al.} 2012, 1

\bibitem[{{The Astropy Collaboration} {et~al.}(2013){The Astropy
  Collaboration}, Robitaille, Tollerud, Greenfield, Droettboom, Bray, Aldcroft,
  Davis, Ginsburg, Price-Whelan, Kerzendorf, Conley, Crighton, Barbary, Muna,
  Ferguson, Grollier, Parikh, Nair, G{\"{u}}nther, Deil, Woillez, Conseil,
  Kramer, Turner, Singer, Fox, Weaver, Zabalza, Edwards, Bostroem, Burke,
  Casey, Crawford, Dencheva, Ely, Jenness, Labrie, Lim, Pierfederici, Pontzen,
  Ptak, Refsdal, Servillat, \& Streicher}]{TheAstropyCollaboration2013}
{The Astropy Collaboration}, Robitaille, T.~P., Tollerud, E.~J., {et~al.} 2013,
  33, 1

\bibitem[{Twicken {et~al.}(2018)Twicken, Catanzarite, Clarke, Girouard,
  Jenkins, Klaus, Li, McCauliff, Seader, Tenenbaum, Wohler, Bryson, Burke,
  Caldwell, Haas, Henze, \& Sanderfer}]{Twicken2018}
Twicken, J.~D., Catanzarite, J.~H., Clarke, B.~D., {et~al.} 2018, Publ. Astron.
  Soc. Pacific, 130, 064502

\bibitem[{van~der Walt {et~al.}(2011)van~der Walt, Colbert, \&
  Varoquaux}]{VanderWalt2011}
van~der Walt, S., Colbert, S.~C., \& Varoquaux, G. 2011, Comput. Sci. Eng., 13,
  22

\end{thebibliography}

\end{document}